\documentclass[useAMS,usenatbib,aas_macros]{mn2e}
\usepackage{aas_macros}
\usepackage{amsmath}
\usepackage{graphicx}
\usepackage{xspace}
\usepackage{amssymb}
\usepackage{booktabs}
\usepackage{float}
\usepackage{hyperref}
\usepackage{color}
\newcommand{\equ}[1]{eq.~(\ref{eq:#1})}

\newcommand{\se}[1]{\S\ref{sec:#1}}
\newcommand{\fig}[1]{Fig.~\ref{fig:#1}}

\newcommand{\Fig}[1]{Figure~\ref{fig:#1}}

\newcommand{\bea}{\begin{eqnarray}}
\newcommand{\eea}{\end{eqnarray}}

\newcommand{\msun}{M_\odot}

\newcommand{\ifm}[1]{\relax\ifmmode#1\else$\mathsurround=0pt #1$\fi}
\newcommand{\kms}{\ifmmode\,{\rm km}\,{\rm s}^{-1}\else km$\,$s$^{-1}$\fi}

\newcommand{\kpc}{\,{\rm kpc}}
\newcommand{\pc}{\,{\rm pc}}
\newcommand{\Gyr}{\,{\rm Gyr}}
\newcommand{\Myr}{\,{\rm Myr}}
\newcommand{\yr}{\,{\rm yr}}

\newcommand{\ltsima}{$\; \buildrel < \over \sim \;$}
\newcommand{\lsim}{\lower.5ex\hbox{\ltsima}}
\newcommand{\gtsima}{$\; \buildrel > \over \sim \;$}
\newcommand{\gsim}{\lower.5ex\hbox{\gtsima}}

\newcommand{\ramses}{{\it RAMSES} }

\usepackage{color}

\begin{document}
\title[Galaxy Evolution: Non-thermal Pressure in ISM]{Galaxy Evolution: Modeling the Role of Non-thermal Pressure in the Interstellar medium}
\author[Y. Birnboim, S. Balberg \& R. Teyssier]{Yuval Birnboim$^\text{1}$, Shmuel Balberg$^\text{1}$, \& Romain Teyssier$^2$\\$^1$Racah Institute of Physics, The Hebrew University, Jerusalem 91904 Israel\\$^2$Institute for Computational Science, University of Zürich, Winterthurerstrasse 190, CH-8057 Zürich, Switzerland\\}
\date{Accepted ---. Received ---; in original ---}

\pagerange{\pageref{firstpage}--\pageref{lastpage}} \pubyear{2014}
\maketitle
\label{firstpage}

\begin{abstract}
  Galaxy evolution depends strongly on the physics of the interstellar
  medium (ISM). Motivated by the need to incorporate the properties of
  the ISM in cosmological simulations we construct a simple method to
  include the contribution of non-thermal components in the
  calculation of pressure of interstellar gas.  In our method we treat
  three non-thermal components - turbulence, magnetic fields and
  cosmic rays - and effectively parametrize their amplitude. We assume
  that the three components settle into a quasi-steady-state that is
  governed by the star formation rate, and calibrate their magnitude
  and density dependence by the observed Radio-FIR correlation,
  relating synchrotron radiation to star formation rates of
  galaxies. We implement our model in single cell numerical simulation
  of a parcel of gas with constant pressure boundary conditions and
  demonstrate its effect and potential. Then, the non-thermal pressure
  model is incorporated into \ramses and hydrodynamic simulations of
  isolated galaxies with and without the non-thermal pressure model
  are presented and studied. Specifically, we demonstrate that the
  inclusion of realistic non-thermal pressure reduces the star
  formation rate by an order of magnitude and increases the gas
  depletion time by as much. We conclude that the non-thermal pressure
  can prolong the star formation epoch and achieve consistency with
  observations without invoking artificially strong stellar feedback.
\end{abstract}


\section{Introduction}
Both star formation and stellar feedback play crucial roles in galaxy
evolution. Star formation leads to stellar feedback, which in turn is
assumed to regulate the star formation rate and prevent galaxies
from turning all their gas into stars over less than a $\Gyr$;
moderate star formation rates are implied from low and high redshift
observations. In contrast, pure hydrodynamic simulations and
semi-analytic models of galaxy formation  tend to predict high gas
densities within galaxies. These densities cause the gas to cool very
efficiently and supersede the density threshold required for star
formation \citep{schmidt59,kennicutt98}. For purely hydrodynamic
simulations, unrealistically
strong stellar feedback is often necessary to regulate the star formation
rate in galaxies \citep{Scannapieco2012}.

Several approaches have been attempted to regulate simulations of star
formation during galaxy evolution.  Within the framework of pure
hydrodynamics, the most basic feedback mechanism is usually thermal
feedback, or injection of some fraction of the supernova energy into
the gas. Since this energy is injected into dense, cold gas it cools
efficiently and typically has small overall effect on galaxy evolution
\citep{Scannapieco2012}.  Momentum feedback is added by explicitly
injecting momentum to the material that surrounds star forming regions
\citep{Navarro1993,springel03_sf,Oppenheimer2006,Dubois2008}. While
the efficiency of such models is slightly better than thermal feedback
\citep{Scannapieco2012}, the ejected gas is almost always highly
supersonic and kinteic energy is converted to thermal radiation very
efficiently through shocks, unless the feedback is injected
effectively over very large volumes \citep{Oppenheimer2006}. Attempts
have also been made to inject the energy into a warm component of a
two-phase gas, effectively delaying the cooling until energy is
transferred from the diffuse gas to a denser gas which immediately
cools \citep{springel03_sf,Governato2007}.  An additional technique
sometime used has been to increase the efficiency of the feedback by
releasing the feedback energy in bursts rather than spread out
\citep{Crain2009}.  However, in all these recipes, the cooling of a parcel
of gas near the plane of the disc still occurs with isobaric boundary
conditions set by the weight of the atmosphere on top of it. Once the
gas cools, it contracts over a crossing time to regain its
pressure. Since the cooling rate scales as $\rho^2$ (where $\rho$ is
the gas density) this leads to runway cooling.  In order to increase
the efficiency of supernova feedback, the cooling of injected energy
is artificially delayed \citep{springel03_sf}, and momentum feedback
efficiency in enhanced by preventing it from interacting with its
immediate, dense environment \citep{Oppenheimer2006}.

In this paper we revisit the conjecture that non-thermal processes
contribute to the total pressure. With this additional pressure, gas
can reach hydrostatic equilibrium with a considerably lower gas
density that naturally predicts lower star formation rate and bypasses
the need for unrealistic supernovae feedback. The non-thermal pressure
does not depend on the temperature of the gas, and the gas cools
isochorically rather than isobarically, further stabilizing the
gas. We develop a simple, easy to use, parametric model which allows
to study (analytically and in simulations) the effect of such
non-thermal components on galaxy formation.

The enhanced star formation problem is closely related to the general
complication of modeling the interstellar medium (ISM) gas. While the
use of a standard, purely thermodynamic equation of state of an ideal
gas is justified outside of galaxies in the IGM, it becomes less
appropriate to use in haloes (haloes of galaxy clusters exhibit
non-negligible magnetic fields) and even more so in the ISM of
galaxies.  This gas is highly multiphased, and consists of cold, warm
and hot gas arranged within atomic and molecular clouds, filaments,
and bubbles \citep{mckee77,ferriere01}. Complicated chemistry and
dynamics, as well as radiation fields at multiple wavelengths affect
the behaviour and interrelation between the different
phases. Moreover, the dynamics of the gas are strongly affected by
non-thermal components, namely turbulence, magnetic fields and cosmic
rays (CR). Stars, through their formation, evolution and destruction
pump energy into the ISM by stirring turbulence, emitting high energy
particles (cosmic rays) and releasing radiation that heats and drives
the gas \citep{deJong1985,bell03_LIR}. Gravitational energy also
powers turbulence \citep{dekel09} and heats the gas \citep[see
however ][]{Hopkins2013}. Turbulence, could, in principle be
accurately followed by pure hydrodynamics. However, modeling
turbulence requires high resolution and realistic driving of the
turbulence which is still an open question \citep{Schmidt2009}. The
detailed modeling of these effects is the subject of intensive ongoing
efforts \citep[e.g.][]{Korpi1999,MacLow2004,Elmegreen2004,
  Dib2006,Robertson2008,Koyama2009,Hopkins2012,Kim2013}. All these
physical phenomena are determind by the relatively small ($\sim\;$pc)
scale of observed giant molecular clouds (GMCs), large eddies of
turbulence \citep{Schmidt2010}, and tangled magnetic fields.

It is prohibitively challenging to include all the aforementioned
effects and small scales in cosmological simulations. Effective
equations of state for star forming gas are thus constructed, directly
pressurizing the thermal component of the gas
\citep{springel03_sf,schaye08}. An equation of state for subgrid
turbulence has been proposed in \citet{maier09}. \citet{Joung2009}
proposed an effective EoS directly related to star formation rates,
and \citet{Braun2012} incorporated turbulent pressure as a sub-grid
model of the various phases motivated by
\citet{mckee77}. \citet{Scannapieco2010} showed that sub-grid models
for supersonic turbulence can have a large affect on dwarf
galaxies. While these attempts artificially pressurize the gas (as we
propose below) they do not prevent the gas from over-cooling
isobarically and still require the unrealistically strong feedback
described above.  Recent attempts
\citep{Salem2014,Booth2013,Hanasz2013} to simulate the effects of
cosmic ray pressure on star formation and winds in galaxies have shown
that they are able to drive significant outflows and could be
efficient in regulating the star formation of star forming
galaxies. \citet{Booth2013,Salem2014} separately implemented a two fluid
approximation for cosmic rays and gas for single-galaxy simulations,
and propagated the CR as a diffusive component with constant diffusion
coefficient. \citet{Hanasz2013} used an MHD code to simulate
anisotropic diffusion that preferentially diffuses CRs along magnetic
fields. In all three implementations the CRs are found to drive winds
in some cases.  These results are encouraging but the simulations are
of single, ideal galaxies and the suggested methods cannot be easily
extended to the evolution phases of the galaxies and cosmological
initial conditions. Their physics is too detailed and the required
resolutions are too fine to be practical in full hydrodynamic
cosmological simulations. To date, no simulations exists that include
all the physical processes which are important for pressurizing the
gas, and no cosmological-scale simulation will likely be able to
simulate all this physics in the forthcoming future.

In view of all these complications, here we follow a simpler,
alternative avenue, and construct an effective EoS that mimics some of
the main observed characteristics of the non-thermal pressure
components. First, we take advantage of the fact that a scale
separation roughly exists between the parsec-scale phenomena discussed
in the previous paragraph and that of typical observed scales for
vertical scale heights of discs which range between $\approx 100 \pc$
for the Milky Way \citep{ferriere01} and quiescent star forming
galaxies to $1\kpc$ for starbursting high redshift galaxies
\citep{tacconi06}. This implies that in the context of galaxy
formation simulations, it should be sufficient to resolve the ISM on
scales much larger than $\sim$1 parsec in order to reproduce galactic
discs with realistic characteristics. Correspondingly we use a coarse
grained, effective modeling of the ISM. Our model bridges the gap
between the parsec and kilo-parsec scales. Second, we complement this
scale separation with an effective equation of state that is
straightforward to use. In principle, one can attempt to apply a
rigorous treatment of the physical processes that constitute the
non-thermal physics as sub-grid models. However, even if realistic
such modules were constructed, there remains the problem of
stipulating physically-consistant initial conditions: how to seed
magnetic fields, how and when cosmic rays are generated and
accelerated, and what drives turbulence and precisely on which
scales. In addition, one has to relate the initial conditions to star
formation which most likely drives these effects. Hence, we opt for a
simple, easy to use, pressure-density relation for the non-thermal EoS
which we develop below. 

A key feature of our implementation is the calibration of the EoS by
the observed relation between the FIR radiation - a star formation
indicator - and the radio radiation, which constrains the joint energy
content of cosmic rays and magnetic fields. This novel approach
provides a quantitative relation between the gas density and the
magnitude of the non-thermal components. Additional physical
assumptions which are required to complete the specific
parametrization of the effective EoS are then limited to factors of
order unity, rather than being arbitrary.

The structure of this paper is as follows. In \se{eos} we describe
possible modifications for the equation of state of the gas that
manifest some important aspects of the non-thermal pressure
components. In \se{toymodel} we demonstrate the effectivenes of such
modified EoS and and the importance of the non-thermal conmponents in
general, using simple calculations of a point (single-cell) model,
focusing on the regaulation of the overall star formation rate.
\se{ramses} describes the incorporation of the model into the
hydrodynamic code \ramses \citep{Teyssier2002} and the setup and
results of isolated galaxy simulations with and without our model. In
\se{summary} we summarize and discuss our results.

\section{Equation of State}
\label{sec:eos} To incorporate non-thermal pressure we need some
typical scale for its density dependence and amplitude. We will use
the observed FIR-radio relation for a typical value of magnetic field
and. The model we describe in this paper assumes that (i) the
non-thermal components are in equilibrium between themselves in the
sense that energy can move quickly between them, and (ii) that this
equilibrium does not depend on the magnitude of the energy. The first
assumption, of strong coupling between the components, is justified
because the timescales for interactions between the components are
eddy turnaround time for the turbulence, alfv\'{e}nic crossing time
for magnetic fields and diffusion times for cosmic rays, all on scales
much smaller than galactic or cosmologic scales. An eddy turnaround
time for a $1\pc$ eddy rotating at a typical ISM speed of $5km/sec$
would take less than $1\Myr$ - much smaller than cosmological
evolution timescales. The alfv\'{e}nic crossing time is even shorter:
the velocity for $B=5\mu G$ and $n=10^{n-3}\text{cm}^{-3}$ is about
$10km/sec.$ The diffusion coefficients for the cosmic rays indicate an
even smaller timescale for equilibrium across a $1\pc.$ With a typical
diffusion coefficient of $D\sim 10^{27}\text{cm}^2 sec^{-1}$ the diffusion
over this scale-length will take $t\sim L^2/D\sim 300\yr$.

We stress that the second assumption is not necessary for the effect
of the non-thermal pressure to be important, and that we use it below
for the sake of simplicity. There are, nonetheless, strong qualitative
arguments which motivate such strong coupling. Compelling physical
arguments can be made in favor of the interrelation between turbulence
energy and tangled magnetic field\footnote{We distinguish between
  ordered magnetic fields which slowly accumulate over the lifetime of
  a galactic discs due to galactic-scale dynamo effect, and tangled
  magnetic fields on scale of a few parsecs and below that is related
  to, and correlates with, the star formation
  \citep{beck96,deavillez05}.Throughout this paper we shall only be
  concerned with the latter.} energy and cosmic rays. Turbulence and
magnetic fields are naturally related since turbulent flow can
increase the energy in magnetic fields (by elongating and wraping the
field lines), while large magnetic fields tend to rearrange and freeze
the material in order to decrease the length of the flux tubes
\citep[see ][for a discussion including the dependence on magnetic
field geometry and turbulence velocity]{Federrath2011} . Either way,
energy is naturally converted from one component to the other.  We
assume that this qualitative argument holds, even though the dominant
coupling process and the precise energy distribution between the
components might vary somewhat \citep[see discussion and references in
][who also assumes equal energies in turbulence and magnetic fields
for starbursting galaxies]{Lacki2013}. Equipartition between magnetic
and cosmic rays should also be natural \citep[][see however
\citet{stepanov12,lacki10a}]{longair94,lisenfeld96,bell03_LIR}. Cosmic
rays are relativistic electrons and protons (most likely accelerated
during supernavae explosions), and travel along magnetic fields which
confines them to the galaxy with some effective diffusion regulated by
the fields. Strong magnetic fields increase cosmic ray energy loses
through synchotron radiation, and reduce the diffusion rate thus
increasing the steady-state density of the cosmic rays in the galaxy.

The simplest manifestation of such an approach is to assume
equipartition between the three non-thermal components, so that the
total energy density is three times that of each separate
component. Equipartition between magnetic fields and cosmic rays
corresponds to a minimum of the total energy of cosmic rays and
magnetic fields for a given measured synchrotron radio emission
\citep{Lacki2013a}.  In cases where the two are not in equipartition
the combined non-thermal pressure due to the two components will be
larger. Finally, the cycle is completed by instabilities in the CR
flows along magnetic fields giving rise to small scale turbulence
\citep{Kulsrud1969}. It is worth mentioning that an equipartition
between these three components are also observed in intra-cluster
medium (ICM) where they each contribute about $5\%$ of the total
pressure, while the bulk of the pressure comes from the thermal
component of X-ray emitting gas. We emphasize that the ``equipartition
anzatz'' is basically just a scaling parameter for the effective
non-thermal EoS; it is not an essential component of our model and
different scalings can be used (see below).

Once the distribution of energy among the non-thermal components is
determined, we can naturally continue to develop an equation of state
for them. While a ``proper'' thermal equation of state relates all the
thermodynamic variables to two independent variables (for example, the
density and internal energy), a non-thermal component is, by nature, a
single parameter function. In the case of magnetic fields, for example,
if a volume of space occupying magnetic fields is compressed, no heat
is generated, and the process can be reversed \footnote{Throughout the
  paper we assume the flux-frozen approximation for magnetic fields
  and neglect magnetic reconnection and ambipolar diffusion, as is
  appropriate for the densities and ionizations of the ISM
  gas.}. Hence, the magnetic pressure and energy are functions of only
one variable (the field magnitude, $B$) which can in turn be related
to only one thermodynamic variable. This is completely analogous to
the equation of state of cold matter, which is commonly used in the
analysis of compact objects.

It is important that the concept of entropy does exist in
multi-component non-thermal system, through the requirement for
equilibrium. We note that the system is not closed, since energy is
constantly pumped in by star formation and leaks out of the system by
cosmic ray diffusion, electromagnetic radiation, reconnections, and
dissipation of turbulence. Hence the entropy of the non-thermal
component can change while generating the equilibrium
configuration. In other words a cold component is not a zero entropy
system (for example, we cannot use the adiabatic relation between the
work done on an element and the internal energy within it). Again, in
analogy with compact objects, this is the basis for deteminning the
composition of high density matter in neutron stars, while allowing
for energy loss through the emission of neutrinos. As mentioned above,
in the absence of a first-princples model for the relation between
magnetic fields, cosmic and turbulence, we simply assume that
equipartition exists between these three components. This
simplification allows us to evaluate the entire non-thermal pressure
based on a relation between one of these components and the star
formation rate, and dictating the energy density in the other two
components by equating it to the first. 

In accordance with this approach, we base our derivation on relations
between magnetic fields and the star formation rate. 
Specifically, we
model the dependence of the magnitude of magnetic field, $B$, on the
star formation rate, $\dot{\rho_*}$ (in mass per unit time per unit
volume) as a power law: $B\propto \dot{\rho_*}^{\alpha_\text{1}}$. Combined
with a \citep{schmidt59} power law relating the star formation rate to
the gas density, $\rho$, i.e., 
\begin{equation}
\dot{\rho_*}=K\rho^\kappa,\label{eq:sfr}
\end{equation}
 we have a
simple power law term of
\begin{equation}\label{eq:B_rho_alphaprime}
B\sim \rho^{\alpha_\text{2}},
\end{equation}
for which $\kappa$, $\alpha_\text{1}$ and $\alpha_\text{2}=\kappa\alpha_\text{1}$ are all
constants, which along with the proportionality factors must be
constrained from observations. Once this assumption has been made, the
total non-thermal volumetric energy arising from these power laws
takes the form: $E_\text{nt}=3 B^2/8\pi\propto\rho^\alpha$,
$\alpha=2\alpha_\text{2}$ with the prefactor of $3$ originating from the
contributions of the $3$ components in equipartition.  We reiterate
that the equipartition is not a necessary assumption for this
model. Any constant distribution between the components is consistent
with the assumptions and can be readily used.  The effective equation
of state can easily be modified for further deviations from
the``equipartition anzatz'', and any non-thermal pressure that is a
monotonic function of density can be incorporated in an analogous
way. Essentially, the only requirement that cannot be simply
generalized is that the non-thermal components are in quasi-steady
state that depends on the gas density alone. That is, that the
non-thermal processes are related, and settle down on timescales which
are short with respect to the evolution of galaxies.

\subsection{Stationary Non-thermal EoS}
\label{sec:ssnteos}
We begin with the simplest model that incorporates the additional
non-thermal pressure components. In this model we assume that the
energy in the non-thermal components is completely determined by the
local instantaneous star formation rate. Stipulating this assumption
the energy in the non-thermal components becomes a simple function of
$\rho$. The function is determined by various physical processes that
contribute to the non-thermal pressure of the ISM, which depend
differently on the density of the gas, but eventually materializes
through equipartition. At this point we will also assume
that the gas is always in appropriate conditions so that star
formation is turned on.

Assuming a power-law dependence of $B$ (Equation
\ref{eq:B_rho_alphaprime}) and a Schmidt law for the star formation,
the non-thermal pressure in equilibrium, $P_\text{nt}^\text{0}(\rho)$,
is: \begin{equation} P_\text{nt}^\text{0}(\rho)=
  A\rho^\alpha,\label{eq:Pnt} \end{equation} where $A$ is a model
dependent proportioanlity factor.  Here, and throughout the paper, we
neglect the order-of-unity differences between pressure and energy
density. Magnetic field's pressure depends on the field's morphology
and drops from $1$ for magnetic field in the disc's plane, to $1/3$
for isotropically tangled field. CR pressure equals $2/3$ of its
energy density for non-relativistic particles. We show below that the
while the detailed analysis changes somewhat, the qualitative effect
of the non-thermal pressure remains unaffected. The total pressure at
each point is the sum of the thermal and non-thermal pressure,
$P_\text{tot}=P_\text{th}+P_\text{nt}^\text{0}$. As an aside we mention that the sound speed
of the gas is striaghtforward to calculate in this effective EoS: For
application in numerical codes, it is useful (for setting the
timesteps according to the Courant conditions, for example) to
calculate the numerical speed of sound of gas:
\begin{align}
c_\text{s}^2&=\left(\frac{\partial P}{\partial \rho}\right)_\text{s}=\frac{\partial
  P_\text{nt}}{\partial \rho}+\left(\frac{\partial
  P_\text{th}}{\partial \rho}\right)_\text{s}\\
&=\alpha\frac{P_\text{nt}}{\rho}+\gamma\frac{P_\text{th}}{\rho}.\nonumber
\end{align}
We note that it is not the physical speed of sound of the
multi-component gas, which depends on the
alfv\'{e}nic velocity and largest eddy velocity in a non-trivial manner.

\subsection{Dynamic Non-thermal EoS}
\label{sec:dnteos}
The EoS described in the previous subsection has the advantage of
being extremely simple to implement since it requires a trivial
addition to the ideal equation of state depending only on the gas
density; There is no need to specifically trace the non-thermal
component. However, it suffers from undesired consequences that arise
from the assumption that the non-thermal pressure traces the local
instantaneuos star formation. This means that if star formation was to
suddenly begin (by passing the a threshold density, for example) or to
suddenly end (perhaps if feedback blowing of the gas transfers it to a
different thermodynamic regime where star formation is extinguished) a
sudden jump in the pressure will follow, and potentially create
spurious shocks and disturbances. In addition, we know from
observations that the magnetic field and cosmic ray vertical scale
heights are considerably larger than those of the gas and star
formation. \citet{ferriere01} estimates the magnetic scale height of
the Galaxy at $\sim 1.4kpc$ based on rotation measures of pulsars
\citep{Inoue1981} and the CR scale height of $\sim 2-4\kpc$ based on
observed abundance of secondary particles such as $^{10}Be$ that are
used to constrain various diffusion models for the Galaxy
\citep{Garcia-Munoz1987,Bloemen1993}. The magnetic field of distant
galaxies is also observed via polarization and radio emission
measurements at a vertical scale of a few kiloparsecs above the disc
plane \citep[see, for example][]{Krause2014}. The existence of
non-thermal pressure based on radio observations and on a discrepancy
between the vertical density and gravity profiles of the Galaxy
perpendicular to the plane at the local neighborhood was advocated by
\citep{boulares90,cox05_ism}. The fact that magnetic fields and cosmic
rays at regions that are far from star forming gas indicate that the
coupling between the non-thermal components and star formation is more
complicated than the simple assumptions in \se{ssnteos}. Specifically,
this suggests that the coupling is not instantaneous, but has a finite
response time as energy convects with the gas or diffuses through
it. Alternatively, there may be additional sources for turbulence,
magnetic fields and CR which are dominant at $\approx 1kpc$ altitudes
above the disc (see, for example, \citet{dekel09} for extra-galactic
driving of turbulence and \citet{Braun2012} for internal ISM
instability-driven turbulence).

To address this complication we introduce another independent variable
into the equation of state so that the amount of non-thermal energy
responds to star formation over a finite time. We set $P_\text{nt}^\text{0}$
(Equation \ref{eq:Pnt}) as the equilibrium value of non-thermal
pressure for a given stellar density, and add a time-dependent form of
the actual non-thermal pressure, $P_\text{nt}(t)$, which approaches
$P_\text{nt}^\text{0}$ through some temporal dependence.
This adjustment expands the stationary EoS and includes a time
integrated function of the star formation rate in the non-thermal
pressure, thus ensuring both temporal and spatial continuity even if
star formation flicks on and off.  Keeping with the spirit of our
model, we do not attempt to describe the physics of the relaxation of
the non-thermal component, and settle for a parametric description. We
note that this formulation also removes the numerical complications
which arise from discontinuities (in the latter sense, this additional
term has a stabilizing effect similar to the von-Neumann artificial
viscosity which was introduced to help integrate over the non-smooth
shock conditions).  We accomplish this by paramerterizing the
non-thermal heating and cooling rate. Since the two must cancel each
other for a steady-state star formation, $P_\text{nt}^\text{0}$ must be an
attractor of $P_\text{nt}(t)$ at a any given mass density: if $P_\text{nt}$ is
too large then there should be a net cooling and vice versa.

Non-thermal heating can be described by: 
\begin{equation}
\mathcal{H}_\text{nt}=f_\text{nt}~\mathsf{\epsilon}_\text{SN}~\eta_\text{SN}~\dot{\rho_*},\label{eq:heating}
\end{equation}
where $\mathsf{\epsilon}_\text{SN}$ is the total energy injected into
the gas per supernova, $\eta_\text{SN}$ is the number of supernovae per
solar mass of stars that are created, and with $f_\text{nt}$ being the
smaller than unity fraction of the supernova energy that ends up in
the non-thermal components. Non-thermal cooling is assumed to be
responding to heating by the following characterization:
\begin{equation}
\Lambda_\text{nt}=f_\text{nt}~\mathsf{\epsilon}_\text{SN}~\eta_\text{SN}
\left(\frac{P_\text{nt}}{P_\text{nt}^\text{0}}\right)^\beta~\dot{\rho_*},\label{eq:cooling1}
\end{equation}
where $\beta$ is a free parameter which essentially controls the
response time of the non-thermal components to changes in the
star-formation rate. The resemblance between the cooling term and the
heating term arises from the requirement that the gas be in
cooling/heating equilibrium at a star formation rate consistent with
observations. Combining these heating and cooling terms the
time-dependent evolution of the non-thermal pressure at a constant gas
density follows the simple form :
\begin{equation}
\dot{P_\text{nt}}=\mathcal{H}_\text{nt}-\Lambda_\text{nt}=f_\text{nt}~\mathsf{\epsilon}_\text{SN}~\eta_\text{SN}
~\dot{\rho_*}\left[1-\left(\frac{P_\text{nt}}{P_\text{nt}^\text{0}}\right)^\beta\right].\label{eq:P_nt_oft}
\end{equation}
Stability requires that:
\begin{equation}\label{eq:stability} \left.\frac{\partial\dot{P_\text{nt}}}{\partial
    P_\text{nt}}\right|_{P_\text{nt}^\text{0}}=-f_\text{nt}~\mathsf{\epsilon}_\text{SN}~\eta_\text{SN}\beta~\dot{\rho_*}<0,
\end{equation}
so that $\beta>0$ ensures that the non-thermal pressure always
approaches its asymptotic value for a steady star formation rate.

The time dependent modification makes it possible to explicitly deal
with a density threshold condition, as observed by
\citet{schmidt59,kennicutt98}. This condition cuts off star formation
completely for gas densities below a critical value, $\rho_\text{c}$. We note
that most numerical codes apply such a threshold (but for considerably
lower threshold densities) also in order to prevent spurious star
formation from occurring outside of galaxies. Below this threshold,
the steady-state non-thermal pressure is expected to vanish. However,
in reality, a non-star forming region can still maintain a steady
state non-thermal pressure due to diffusion processes from neighboring
regions \citep{Joung2009,Scannapieco2012}, or to various other sources
\citep{Braun2012,dekel09}. We partly account for that within our
``single-cell'' framework by setting a finite decay rate for
non-star-forming regions.  In \equ{cooling1}, we cannot set
$P_\text{nt}^\text{0}=0$ for $\rho\leq \rho_\text{c}$, since the cooling rate then
becomes ill defined.  We remedy this by formulating $P_\text{nt}^\text{0}$ as a
function of $\rho$ (rather than $\dot{\rho_*}$) and redefining
(\equ{Pnt}) as follows:

\begin{align}
\label{eq:P_nt_0_new}
P_\text{nt}^\text{0} = A\rho^{\alpha '}&\text{ for } \rho>\rho_\text{c}\\
P_\text{nt}^\text{0} = A\rho_\text{c}^{\alpha '} &\text{ for } \rho<\rho_\text{c}.\nonumber
\end{align}
Using \equ{heating} and \equ{cooling1} now assures that the heating
turns off when no star formation occurs, and the cooling can proceed
as the non-thermal pressure asymptotically approaches $0$. Combining
\equ{P_nt_oft} with \equ{sfr} then yields:
\begin{equation}
\dot{P_\text{nt}}=\mathcal{H}_\text{nt}-\Lambda_\text{nt}=f_\text{nt}~\mathsf{\epsilon}_\text{SN}~\eta_\text{SN}\left[
~\dot{\rho_*}-K\rho^\kappa\left(\frac{P_\text{nt}}{P_\text{nt}^\text{0}}\right)^\beta\right].\label{eq:P_nt_dot}
\end{equation}

There is a noteworthy simplification in our model. In order to achieve
coarse-grained steady state of ISM gas two conditions must be met
simultaneously: the total pressure must balance the external pressure,
and the cooling must balance the heating. For purely thermal pressure,
these two equations are solved by varying two parameters - the density
and temperature of the gas. For gas with purely non-thermal pressure
components of the type proposed in this work (which is a reasonable
approximation for many external pressures, see \fig{extp}), the
pressure is a function of density alone, and the pressure equilibrium
and heating/cooling equilibrium generally do not have a simultaneous
solution. We bypass this by relating the heating to the cooling in
such a way that the observed relation is always achieved. More
advanced models which include physically motivated cooling and heating
will introduce more dynamic parameters, allowing the gas to reach
steady state more naturally.

\section{Dynamic Behaviour of the Non-Thermal EoS: A Quantitative
  Model}
\label{sec:toymodel}
We now demonstrate the proporties and applicability of our effective
EoS for non-thermal components with a point (zero-dimensional) model
of the ISM. In this model we evolve the conditions of a parcel of gas
with isobaric boundary conditions, solving both thermal and
non-thermal pressure components, in accordance with the models
described in \se{eos}.

\subsection{Model Parameters}

A quantitative implementation of our EoS requires the speficiation of
the model's free parameters. For the stationary non-thermal pressure,
these are the proportionality coeffecient and power which relate gas
density to the pressure in magnetic fields
($P_\text{nt}^\text{0}=A\rho^\alpha$). Even after applying our hypothesis of
equaipartition among the non-thermal components, current uncertainties
regarding the magnitude of magnetic fields in early galaxies, are quite large.
In essence, $(A,\alpha)$ may be treated as free
parameters. Some indication can, however, be gained from observed
relations between star formation rates and synchrotron radiation \citep{Kennicutt83}. 
We chose to use the fits from equation A11 of \citet{lacki10b} of the
form: 
\begin{equation}
B=B_\text{0}\Sigma_\text{gas}^a h^{-a},
\end{equation}
where $\Sigma_\text{gas}$ and $h$ are galactic gas column densities and
galactic scale-heights of galaxies. These two parameters are fitted to
observations using a one zone model for galaxies including the cosmic
ray specra of primary and secondary rays, tracing self-consistently
generation and evolution with effective diffusion coefficients
\citep{lacki10a,lacki10b,Lacki2013}. The power law coefficient $a$,
and normalization $B_\text{0}$, are observationally constrained by the FRC
\citep[far IR - radio correlation][]{Condon1991,Yun2001} and by local
measurements of CR and radio observations at $1.4GHz$ \citep[see ][and
reference within]{lacki10a}. A model for the turbulent amplification
rate of magnetic fields by star formation that recovers the FRC and
predicts its breakdown was also suggested by
\citet{Schleicher2013}. Stipulating a typical vertical height for the
magnetic fields of $1\kpc$ \citep{cox05_ism} one finds for the two fits
suggested by \citet{lacki10a} \footnote{We note that these
  scale-heights are higher than the ones used by
  \citet{lacki10a}. Using a smaller scale-height would result in
  higher magnetic fields and even higher non-thermal pressure.}:
\begin{align}
  B&=6.65\left(\frac{\rho}{10^{-24}\text{gr}~\text{cm}^{-3}}\right)^{0.5}\mu G,\\
  B&=6.85\left(\frac{\rho}{10^{-24}\text{gr}~\text{cm}^{-3}}\right)^{0.6}\mu
  G.\nonumber
\end{align}
These values are consistent with with observed Milky Way values
\cite{ferriere01,cox05_ism,beck09} and with theoretical predictions
\citep{lisenfeld96}.  We convert those relations to non-thermal energy
according to $P_\text{nt}=3~\frac{B^2}{8\pi}$ (again, recalling that the
factor of 3 arises from the equipartition assumption) and find two
similar (but not identical) realizations for the non-thermal EoS:
\begin{align}
  P_\text{nt}^\text{0}(\rho)=5.3\times
  10^{-12}\left(\frac{\rho}{10^{-24}\text{gr}~\text{cm}^{-3}}\right)
  \text{erg}~\text{cm}^{-3}\label{eq:powerlaw1}\\
  P_\text{nt}^\text{0}(\rho)=5.6\times
  10^{-12}\left(\frac{\rho}{10^{-24}\text{gr}~\text{cm}^{-3}}\right)^{1.2}
  \text{erg}~\text{cm}^{-3}.\label{eq:powerlaw2}
\end{align}
According to \citet{lacki10a} these fits reproduce the FIR-radio
relations of galaxies equally well. One could consider a case when the
magnetic pressure scales linearly with density (corresponding to the
first fit), or to the star formation (corresponding to a fit with
$B\sim \rho^{0.75},$ or $B^2\sim \rho_*\sim \rho^{1.5}$) - but this parameter is not favored
by the more detailed model there. As we point below, in our model, the
$B\sim rho^{0.5}$ fit is singular in the sense that its evolution can
never derail it from static equilibrium once it has achieved. For this
reason, it is worthwhile to keep both fits at this stage. 

It is encouraging to note that these values are in rough agreement
with the values needed to support the weight of the gas at the plane
of the Galactic disc against its self gravity \citep{cox05_ism}.

Other tracers of star formation could in principle be used to
calibrate and constrain the non-thermal pressure terms. X-rays are a
good tracer for star formation as young OB stars emit X-rays
\citep{ranalli03,Mineo14} and at high redshifts
\citep{Vattakunnel12}. However, X-rays are converted by neutral gas
efficiently into UV and optical, and are never important as radiation
pressure. Further, X-rays are not a direct tracer for other sources
of pressure of in the gas such as the amplitude of magnetic fields,
turbulence or CR. Measurements of turbulence by emission line
broadening could, in principle, be used to calibrate the relation
between star formation rate and the non-thermal components. However,
such measurements inevitably probe only some of the ISM gas phases, and
measurements of high-redshift turbulence \citep{forster06}
are few and might be driven by infalling gas rather than by star
formation \citep{dekel09}. Alternative direct measurements of
magnetic fields either trace the large scale magnetic fields of
galaxies (polarization measurements) or the total line-of-sight
magnetic fields (Faraday rotation measures) and are harder to
correlate to total star formation than the synchrotron radiation used
in this work.

For the dynamic non-thermal EoS several additional parameters are
required to define equations \equ{heating} and \equ{cooling1}. The
supernova energy $\mathsf{\epsilon}_\text{SN}$ and the supernova rate
$\eta_\text{SN}$ can by taken from standard theories, but we do need to
specify the parameters which control non-thermal heating and cooling,
$f_\text{nt}$ and $\beta$. The fraction $f_\text{nt}$ sets the fraction of
supernovae energy invested as non-thermal energy, while $\beta$ sets
the power governing the rate at which the non-thermal energy reaches
its equilibrium values. These values are numerical by nature and
should be set to allow the non-thermal energy to achieve equilibrium,
while smoothing over pressure jumps arising from abrupt changes in the
star formation rates. As we show below, even for a low $f_\text{nt}$ of
$0.1,$ 
the relaxation times for
a wide choice of $\beta$ are shorter than a $\Myr.$ This value
indicates that for smooth galactic histories the calculated state of
the gas should be close to the asymptotic conditions constrained by
observations.

\subsection{The Single Cell Isobaric Model}
We incorporate our model for non-thermal pressure in a single cell
model by tracing the evolution of gas under isobaric boundary
conditions. This represents a simplified behaviour of a single
hydrodynamic cell embedded within a galaxy that evolves slowly and
supports this cell with nearly constant external pressure. For the
general, dynamic case we solve the ordinary differential equations for
the thermal internal energy of the gas and for the non-thermal
pressure:
\begin{eqnarray}
  \dot{e}_\text{th}=f_\text{th}\epsilon_\text{SN}\eta_\text{SN}\frac{\dot{\rho}_*}{\rho}-\Lambda_\text{t}(\rho,T)+P_\text{th}\frac{\dot{\rho}}{\rho^2},\label{eq:edot}\\
  \dot{P}_\text{nt}=f_\text{nt}~\mathsf{\epsilon}_\text{SN}~\eta_\text{SN}\left[
    ~\dot{\rho_*}-K\rho^\kappa\left(\frac{P_\text{nt}}{P_\text{nt}^\text{0}}\right)^\beta\right]+P_\text{nt}\frac{\dot{\rho}}{\rho}\label{eq:pntdot}.
\end{eqnarray}
The last term in the right hand side of both equations is the
contribution of the density change, $\dot{\rho}$, (a $PdV$ term for
the energy equation).
The density is derived self-consistently by requiring that
\begin{equation}
P_\text{nt}\left(\rho\right)+P_\text{th}\left(\rho,e_\text{th}\right)=P_\text{ext}.
\end{equation}

We complement our model with an appropriate paramerterization of the
star formation rate and the gas heating and cooling functions. Star
formation is modeled with a Schmidt law corresponding to a convention
of $e_\text{eff}=5\%$ of the gas into stars every dynamic free-fall time of
the gas:
\begin{equation}
\dot{\rho_*}=e_\text{eff}\frac{rho}{t_\text{ff}}=e_\text{eff}\left(\frac{32G}{3\pi}\right)^{1/2}\rho^{3/2}.
\end{equation}
For the sake of simplicity we begin with this star-formation rate with
no density cutoff (we examine the implications of such a cutoff in section \ref{sec:cutoff}).  The equation of state for the ideal
gas implies $P_\text{th}=\left(\gamma-1\right)\rho e_\text{th}.$ In this work we
allow the gas to cool according to the rates $\Lambda_\text{t}(\rho,T)$ from
CLOUDY \citep[version 96b4][]{Ferland1998} by interpolating from
tables described in \citet{kravtsov03}. The cooling and heating of the
gas includes Compton heating and cooling, redshift
dependent UV heating and atomic and molecular cooling. The tables
provide the total cooling and heating and particle number as a
function of the redshift, metalicity, density and temperature. The
temperature is related to the internal energy by integrating over the
particle number dependent heat capacity as the number of particles
changes by a factor of a few at recombination and at molecular
formation. For supernova heating we assume that one supernova occurs
for every $160\msun$ of stars formed ($\eta_\text{SN}=1/160\msun$; see
\equ{heating}), corresponding to a Salpeter IMF between $0.1$ and
$100\msun$ and supernovae occurring above $8\msun$ \citep{dobbs11}. We
use a standard value for the average total energy released per
supernova, ${\epsilon}_\text{SN}=10^{51}\text{erg}$.  In most of the simulations
described below we impose external pressure boundary conditions of
$10^{-12}\text{erg}~\text{cm}^{-3}$ in rough correspondence with observed conditions
in the plane of the Galactic disc \citep{cox05_ism}. 

For completeness we list the various definitions and default values
for the coefficients in our equations in table \ref{tab:values}.

\begin{table*}
\caption{Parameters and values of the isobaric gas evolution calculations}
\small
\begin{tabular}{ r r r | l }
\hline
\hline
Parameter & Units & Value & Definition \\
\cline{1-4}\\
\multicolumn{4}{c}{Standard Parameters}\\
\cline{1-4}\\
$\gamma$ & & 5/3 & adiabatic constant \\ 
$z$ & & 0 & redshift \\
$Z$ & $Z_\odot$ & 1 & metalicity \\
$\epsilon_\text{sfr}$ &  & 0.05 & star formation efficiency \\
$\eta_\text{SN}$ &  & 1/160 & supernova per stellar mass formed \\
$\epsilon_\text{SN}$ & erg & $10^{51}$ & supernova energy \\
$f_\text{th}$ &  & 0.1 & fraction of energy injected to thermal component\\
\cline{1-4}\\
\multicolumn{4}{c}{Non-thermal pressure}\\
\cline{1-4}\\

$f_\text{nt}$ &  & 0.2 & fraction of energy injected to thermal component \\
$\alpha$ &  &  & density power law coefficient of non-thermal pressure (\equ{Pnt})\\
$A$ & see \equ{Pnt} &  & normalization of non-thermal pressure \\
$\beta$ &  &  & cooling behaviour (\equ{cooling1}\\
\cline{1-4}\\
\multicolumn{4}{c}{Dynamics of simulations}\\
\cline{1-4}\\
$P_\text{ext}$ & $\text{erg} \, \text{cm}^{-3}$ & $10^{-12}$ & External pressure \\
$P_\text{nt}^\text{i}$ & $\text{erg} \, \text{cm}^{-3}$  &  & initial non thermal pressure \\
$T^i$ & K & $10^5$ & initial gas temperature \\
\hline
\hline
\end{tabular}
  \label{tab:values}
\end{table*}

\subsection{Evolution of Gas with Purely Thermal Pressure}
\label{sec:thermal}

\begin{figure}
\begin{center}
\includegraphics[width=3.3in]{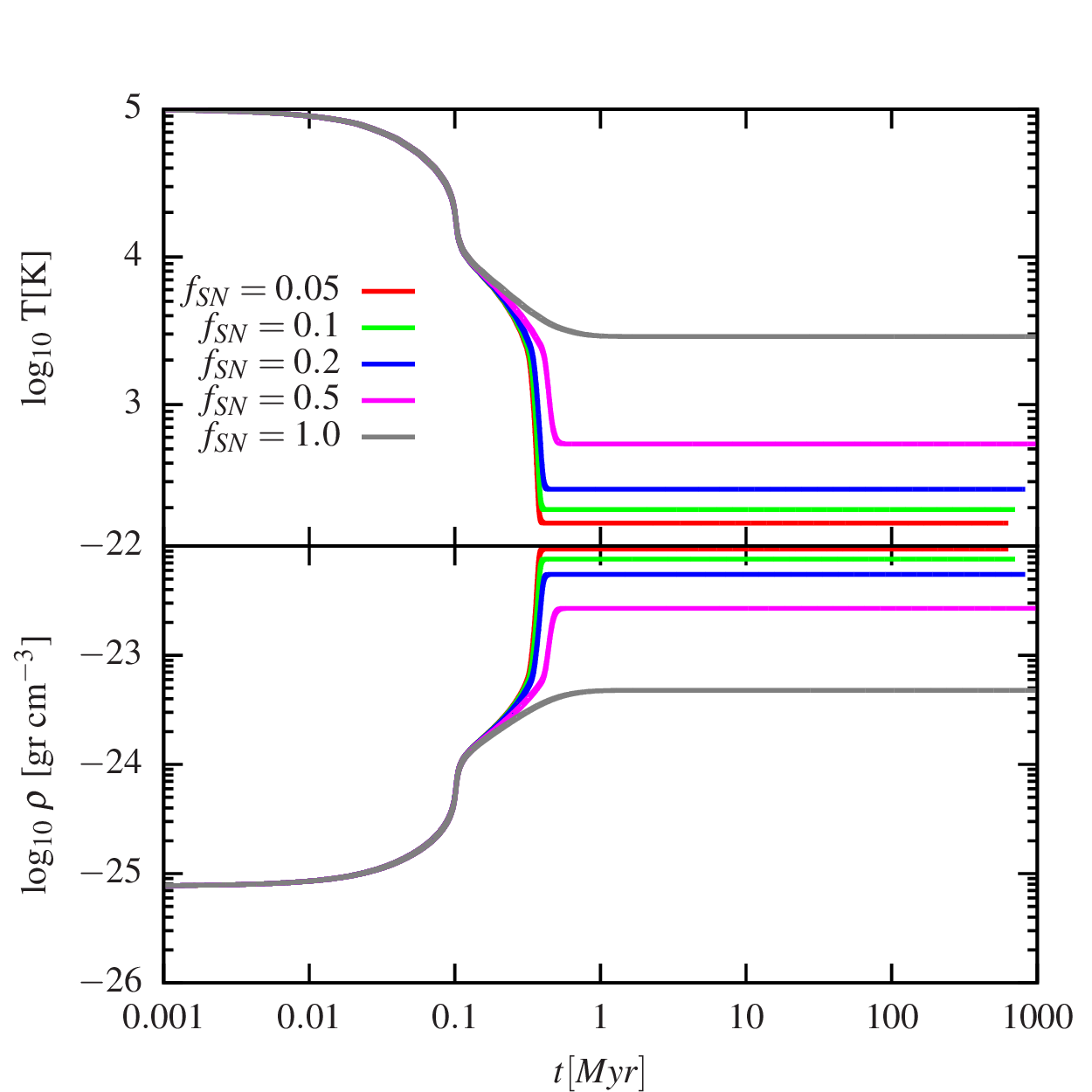}
\caption{\label{fig:thermal_trho} Time evolution of the temperature ({\it top panel})
  and density ({\it bottom panel}) of the thermal-pressure-only models for varying
  supernovae efficiencies. The pressure boundary conditions is
  $10^{-12}\text{dyn}~\text{cm}^{-2}$ and the initial temperature of the gas is
  $10^5$\,K. }
\end{center}
\end{figure}
\begin{figure}
\begin{center}
\includegraphics[width=3.3in]{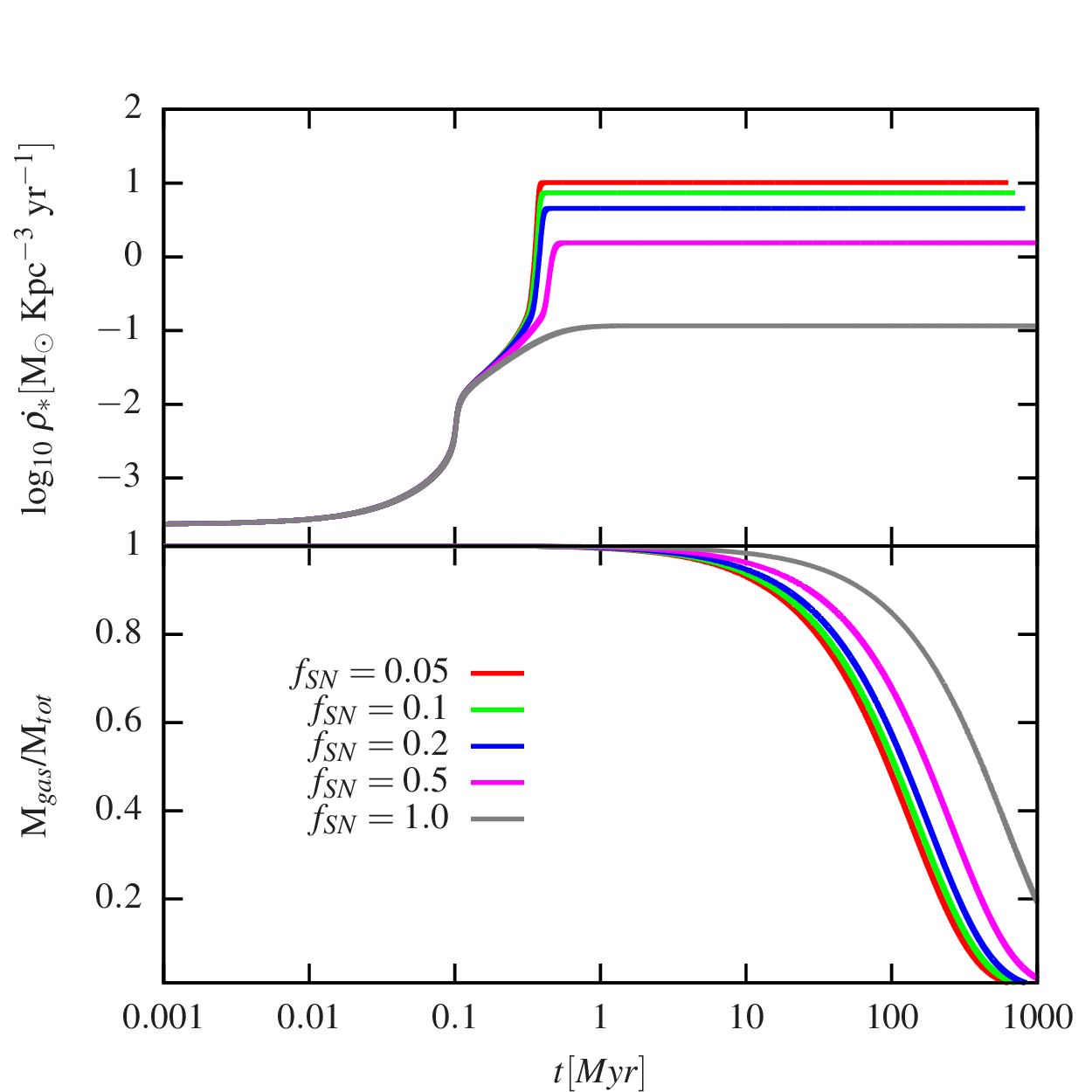}
\caption{\label{fig:thermal_rhodot} Time evolution of the star
  formation rates and the depletion of gas into stars
  ($M_\text{gas}/M_\text{tot}$) for the model calculated in 
  \Fig{thermal_trho} . }
\end{center}
\end{figure}
 
We start by demonstrating the properities of the single-cell
simulation when only thermal pressure exists. We set the initial
conditions with a temperature of $T_\text{i}=10^5$\,K, which corresponds to a
density of $\approx 10^{-25}\text{gr}~\text{cm}^{-3}.$ \Fig{thermal_trho}
demonstrates how cooling of the gas causes the temperature of the gas
to decrease ({\it top panel}), and, correspondingly the isobaric
boundary condition forces the density to scale as $1/T,$ forcing a
density increase ({\it bottom panel}). The bump at $\approx 0.1\Myr$
corresponds to the steep decrease in the cooling function at $T=10^4$\,K
as gas becomes neutral and collisional excitation of lines becomes
unimportant beyond this point.
  
Greater densities enhance the star formation rate, as well as the
resulting supernova feedback, and once the supernova feedback power
balances the cooling rate the density and temperature of the gas
become constant and the gas is converted into stars at a constant
rate. This is can be seen clearly in \fig{thermal_rhodot}, which
depicts the specific star formation rate and the depletion of gas into
stars.  This depletion (\fig{thermal_rhodot}, {\it bottom panel}) is
calculated by noting that the mass in stars evolves as:
\begin{equation}
M_\text{tot}=M_\text{gas}+M_*=M_\text{gas}+\int\dot{\rho_*}Vdt=\text{const},
\end{equation}
where $M_\text{tot},M_\text{gas}$ and $M_*$ are the total, gas and stellar mass
in our volume element, and $V$ and $\rho$ the time-dependent volume
and density of the element.  In the single cell model $M_\text{tot}$ is
fixed. As gas is converted to stars the volume adjusts itself so the
pressure corresponds to the external boundary condition. This single
cell assumption is self-consistently addressed in the full
hydrodynamic implementation shown below (\se{ramses}). Initially,
$M_\text{tot}=M_\text{gas}=V_\text{0}\rho_\text{0}$ with $V_\text{0}$ and $\rho_\text{0}$ the initial volume
and gas density, respectively.
\begin{equation}
\frac{V}{V_\text{0}}=\frac{M_\text{tot}-M_*}{V_\text{0}~\rho}=\frac{\rho_\text{0}}{\rho}-\frac{1}{\rho}\int\dot{\rho_*}\frac{V}{V_\text{0}}dt,\label{eq:vv0}
\end{equation}
is an integral equation that can be evolved in time. The depletion
of gas is then shown as:
\begin{equation}
  \frac{M_\text{gas}}{M_\text{tot}}=1-\frac{M_*}{M_\text{tot}}=1-\frac{1}{\rho_\text{0}}\int
  \dot{\rho_*}\frac{V}{V_\text{0}}dt.\label{eq:depletion}
\end{equation} 
As is to be expected, once the density levels off at an equilibrium
value, so does the specific star formation rate (the volume of the
element continues to decrease over time in order to maintain a
constant gas density with a decreasing mass).

Our key observation is that without non-thermal components the gas
achieves an equilibrium between the heating and the cooling after less
than $1Myr$ and then converts most of the gas into stars quickly after
that. About $50\%$ of the gas is depleted during the first $100\Myr$
for low supernova efficiencies, and even when assuming perfect
($f_\text{SN}=1$) supernovae effeciencies, star formation has comsumed over
one half of the gas by $400\Myr$. Absolute efficiency is certainly
non-physical, since in reality most of the supernova energy gets
converted into radiation that escapes the galaxy without contributing
to pressure support of the gas. In any case, we conclude that in a
thermal-pressure-only model, boundary conditions corresponding to the
pressure in the mid-plane of the Milky Way leads to gas being
converted into stars over a few $100\Myr$ regardless of the efficiency
of the thermal feedback.

\begin{figure}
\begin{center}
\includegraphics[width=3.3in]{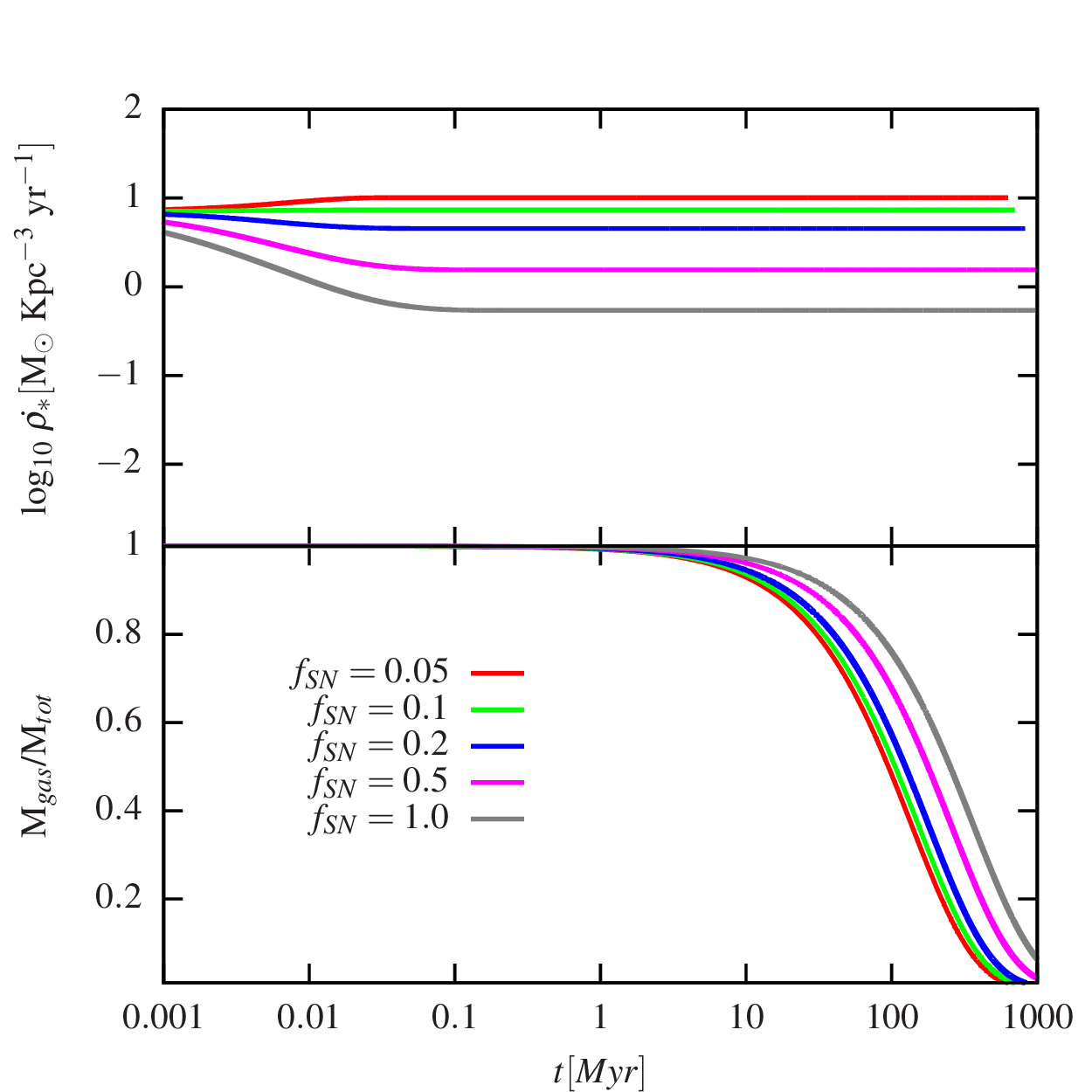}
\caption{\label{fig:thermal2_rhodot} 
Same as \Fig{thermal_rhodot}, but when the initial gas temperature is $200$\,K.
}
\end{center}
\end{figure}

It is also noteworthy that since the gas achieves rate equilibrium during the first $\Myr$, the
initial conditions of the gas do not affect the depletion time. In
\Fig{thermal2_rhodot} we repeat the exercise with an initial
temperature for the gas of $200$\,K, and the results are virtually
unchanged, except that cooling/heating equilibrium is achieved after
as little as $0.1\Myr$. We note that the unphysical $f_\text{SN}=1$ case
reaches a different equilibrium point that exists on the molecular
cooling branch at a lower temperature and leads to even faster gas
depletion.

\subsection{Evolution of Gas with Thermal and Non-Thermal Pressure}
We now proceed to examine the behaviour of a parcel of gas with similar
boundary conditions as in \se{thermal}, but with additional
non-thermal components, evolved according to \equ{pntdot}.

\begin{figure}
\begin{center}
\includegraphics[width=3.3in]{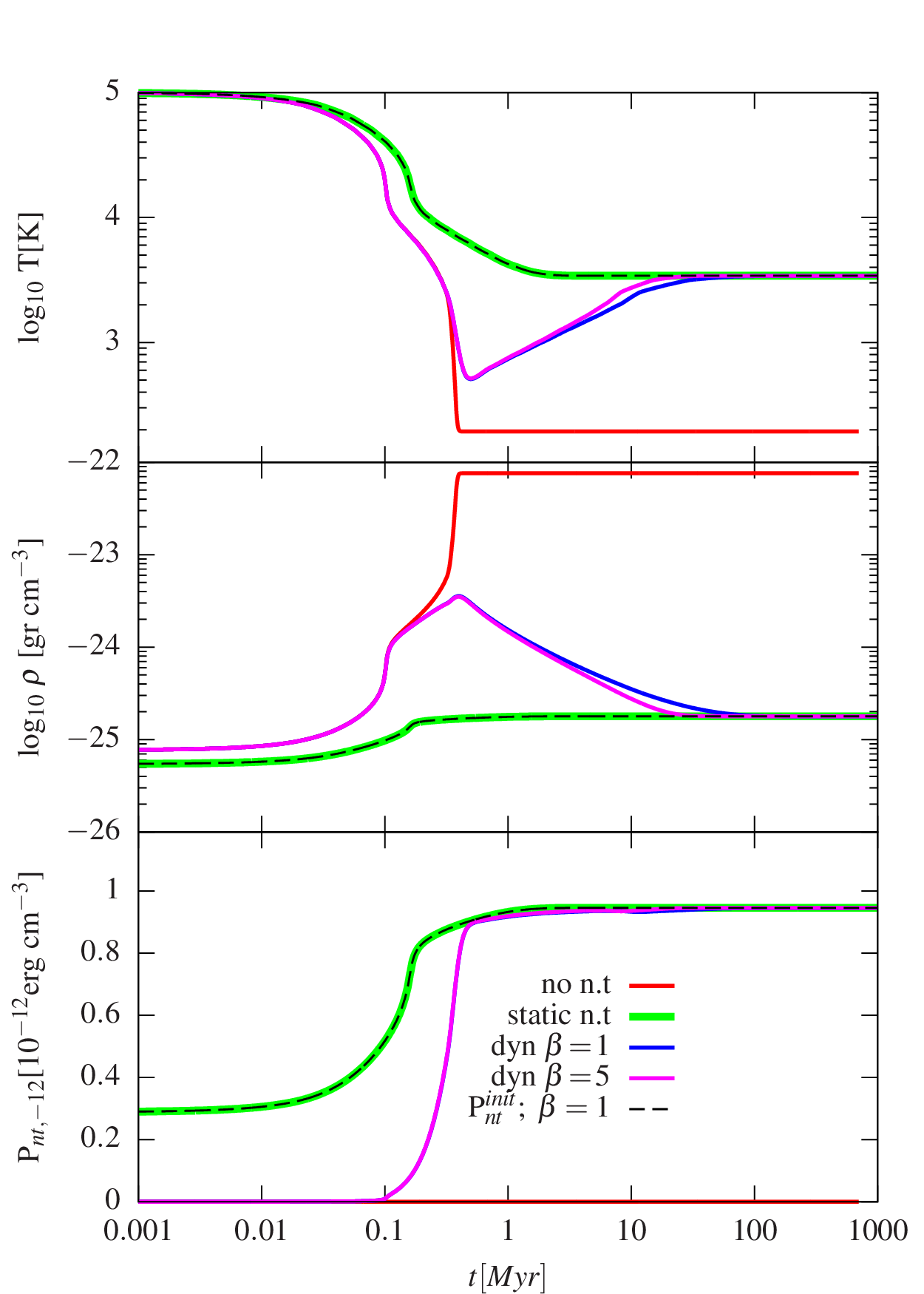}
\caption{\label{fig:ism_trho} Time evolution of the temperature (top
  panel) density (middle panel) and non-thermal pressure (bottom
  panel) of the gas. Curves correspond to no non-thermal pressure
  (red), the steady-state non-thermal pressure (green), and dynamic
  non-thermal pressure (blue, cyan and gray); see text for detail - but note that in this case the green and gray lines overlap completely. In
  all calculations the pressure boundary conditions is
  $10^{-12}\text{dyn}~\text{cm}^{-2}$ and the initial temperature of the gas is
  $10^5$\,K.}
\end{center}
\end{figure}

\begin{figure}
\begin{center}
\includegraphics[width=3.3in]{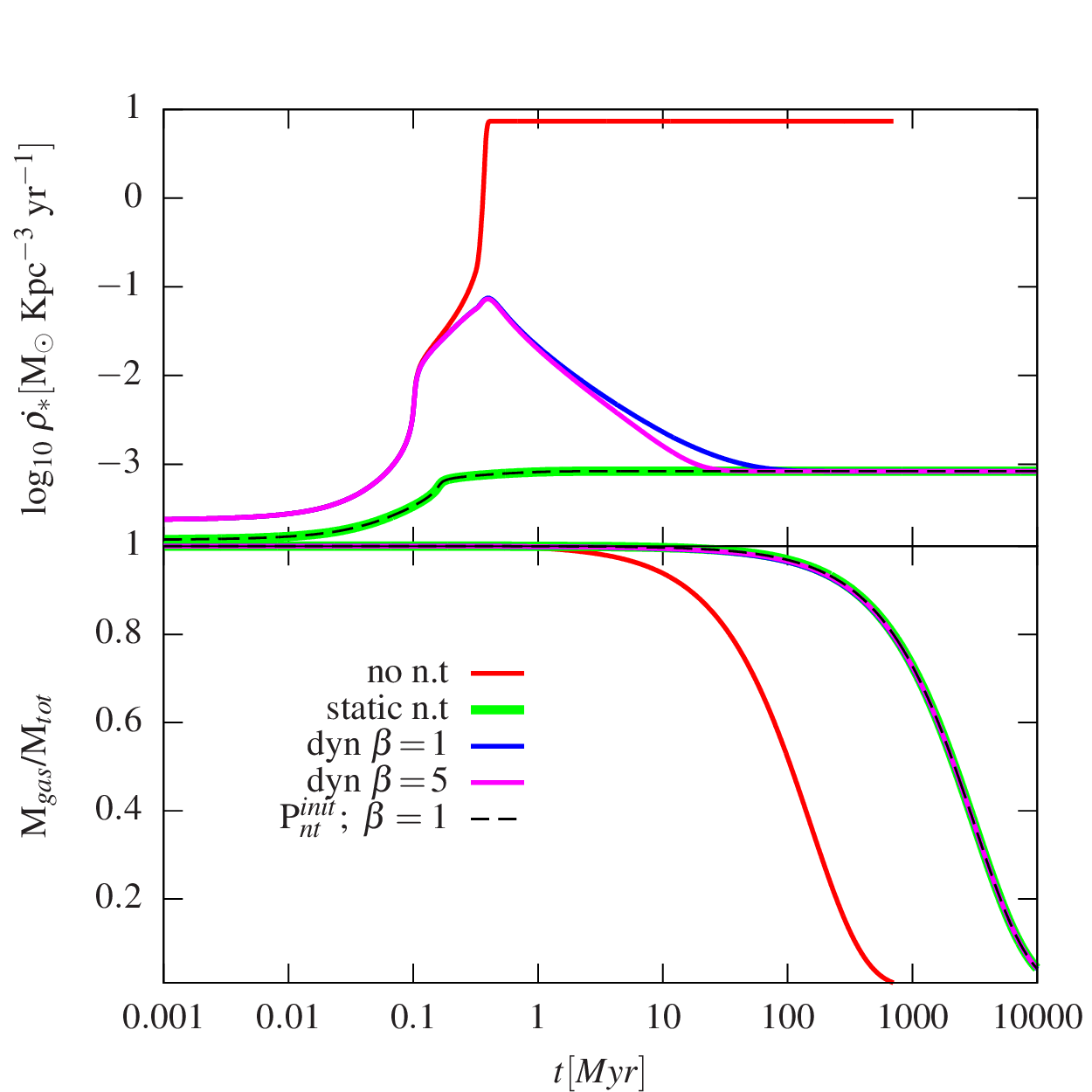}
\caption{\label{fig:ism_rhodot} Time evolution of the star formation
  rates and the depletion of gas into stars ($M_\text{gas}/M_\text{tot}$) for
  the models calculated in \Fig{ism_trho} . }
\end{center}
\end{figure}

\Fig{ism_trho} shows the evolution of a parcel of gas in terms of
temperature, density and non-thermal pressure, again with a pressure
boundary condition of $10^{-12}\text{dyn}~\text{cm}^{-2}$ and and initial
temperature of $10^5$\,K. For the non-thermal pressure we use the
parametrization described in \equ{powerlaw1}. \fig{ism_rhodot}
describes the evolution of the specific star formation rate and gas
depletion for the same model. In all the simulations here the thermal
supernova feedback is turned on with efficiency $f_\text{SN}=0.1$ as
described in \se{thermal}, and the fraction of supernova energy that
is injected into the non-thermal component here is $f_\text{nt}=0.2.$.

The green line shows the stationary ($P_\text{nt}=P_\text{nt}^\text{0}$) non-thermal
EoS described in \se{ssnteos} (calculated by replacing \equ{pntdot}
with \equ{Pnt}), and the blue, cyan and gray lines are for the dynamic
non-thermal EoS (\se{dnteos}) with the relaxation power laws of
$\beta=1$ and $5$ as indicated on the plot. The blue and cyan lines
correspond to models where we arbitrarily set a zero initial
non-thermal pressure, $P_\text{nt}^\text{init}=0.$ This initial condition
results in initial density of $\approx 10^{-25}\text{gr}~\text{cm}^{-3}$ as for the
thermal case, whereas the gray line corresponds to an initial
$P_\text{nt}^\text{init}$ which is in its steady-state value for an initial
density, $\approx 6\times 10^{-26}\text{gr}~\text{cm}^{-3}$. Note that this value is
only slightly below the initial value when non-thermal pressure is
neglected.

We note that the green and gray lines are identical: the power-law in
\equ{powerlaw1} is $\alpha=1,$ and when the gas is in equilibrium
($P_\text{nt}=P_\text{nt}^\text{0}$) its evolution according to \equ{pntdot} is just
$\dot{P}_\text{nt}=P_\text{nt}\frac{\dot{\rho}}{\rho}=A\dot{\rho}$ so it cannot
evolve away from equilibrium once initially achieved. We shall show
below that the static evolution deviates from equilibrium initial
conditions case when $\alpha\ne 1$ (Equation \ref{eq:powerlaw2}).

The two figures clearly demonstrate the distinct effect that
non-thermal pressure has on the simulation. For the initial conditions
we set, the gas is initially supported (at least in part) by thermal
pressure. As the thermal energy is radiated away, temperature drops
and the gas contracts, increasing the star formation rate. However,
the inclusion of non-thermal pressure removes the relation of $\rho
T\propto P_\text{tot}$, and introduces another degree of freedom. The gas can
then cool without a dramatic density increase, and so cooling does not
necessarily lead to enhanced star formation. In the calculations with
the dynamic EoS the non-thermal component adjusts (increases) until a
new stationary equilibrium is reached. This equilibrium consists of a
balance between supernovae feedback and cooling both for the gas and
the non-thermal energy components, each separately. Note that in all
the simulations presented here the gas settles into this steady state
in a few $\Myr$.

The shape of the the curves found with the dynamical EoS also deserves
some elaboration. Since the asymptotic non-thermal pressure is similar
in all these cases, all trajectories with non-thermal pressure
converge to the same values. As gas contracts, its star formation and
supernova rate increases, and, for the dynamic non-thermal EoS, it
takes some time for the non-thermal reservoir to fill. During this
time the gas is actually under-pressurized with respect to its
asymptotic values and the density is larger than its final value. This
overshoot is readily seen in the temperature and density of the gas of
\fig{ism_trho} and in the specific star formation rate of
\fig{ism_rhodot}. The {\it bottom panel} of \fig{ism_trho} shows the
gradual and monotonic increase of $P_\text{nt}$. The timescale for
converging to the asymptotic value is set primarily by $f_\text{nt},$ and
slightly depends on the relaxation power law $\beta.$ In all the runs
here the gas quickly settles into a steady state for which the cooling
is balanced by the thermal feedback and the total pressure is divided
between the thermal component and the non-thermal component.

The distinct effect of non-thermal pressure is easily seen by
comparing the evolution in all of these calculations to the case in
which non-thermal components are neglected, similar to
\fig{thermal_trho} and \fig{thermal_rhodot} (shown for reference in
\Fig{ism_trho} and \Fig{ism_rhodot} in red curves). The outstanding
feature is the dramatic difference in the asymptotic equilibrium
between the two cases: the additional source of pressure allows the
gas to cool without a dramatic density increase. Hence, the
equilibrium density is some $400$ times lower for the non-thermal
case, and the temperature is $10$ times higher. These different
conditions lead to very different star formation rates and depletion
times as can be observed in \fig{ism_rhodot}. Once the non-thermal
component is included, the equlibrium star formation rate is four
order of magnitudes lower (corresponding to the Schmidt law used here
that indicates $\dot{\rho_*}\propto\rho^{1.5}$).  Accordingly, the
mass depletion time for the non-thermal EoS gas increases to $\approx
2\Gyr$ as opposed to about $100\Myr$ when only the thermal component
in the pressure is included.  This increase in depletion times is
related to the lower asymptotic density for this case. Since the
density approaches its asymptotic value much faster then the depletion
time, most of the gas depletion occurs at the equilibrium
density. Hence the gas depletion time is $\tau_*\approx
M_\text{gas}/(V\dot{\rho}_*)=V\rho/(V\dot{\rho}_*)=\rho/\dot{\rho}_*$. For
the Kennicutt Schmidt relation we use here, this leads to a
$\tau_*\sim \rho^{-1/2}$ relation, so reducing the density by a factor
of $400$ leads to a twenty fold increase of the depletion time.

We also note the value of $\beta$ has a minor impact on the
relaxation time scale in the dynamic
models, which is ten to a few tens of Myr (as is to be expected, the
model with $\beta=5$ has a shorter relaxation time than the one with
$\beta=1$).

\begin{figure}
  \begin{center}
    \includegraphics[width=3.3in]{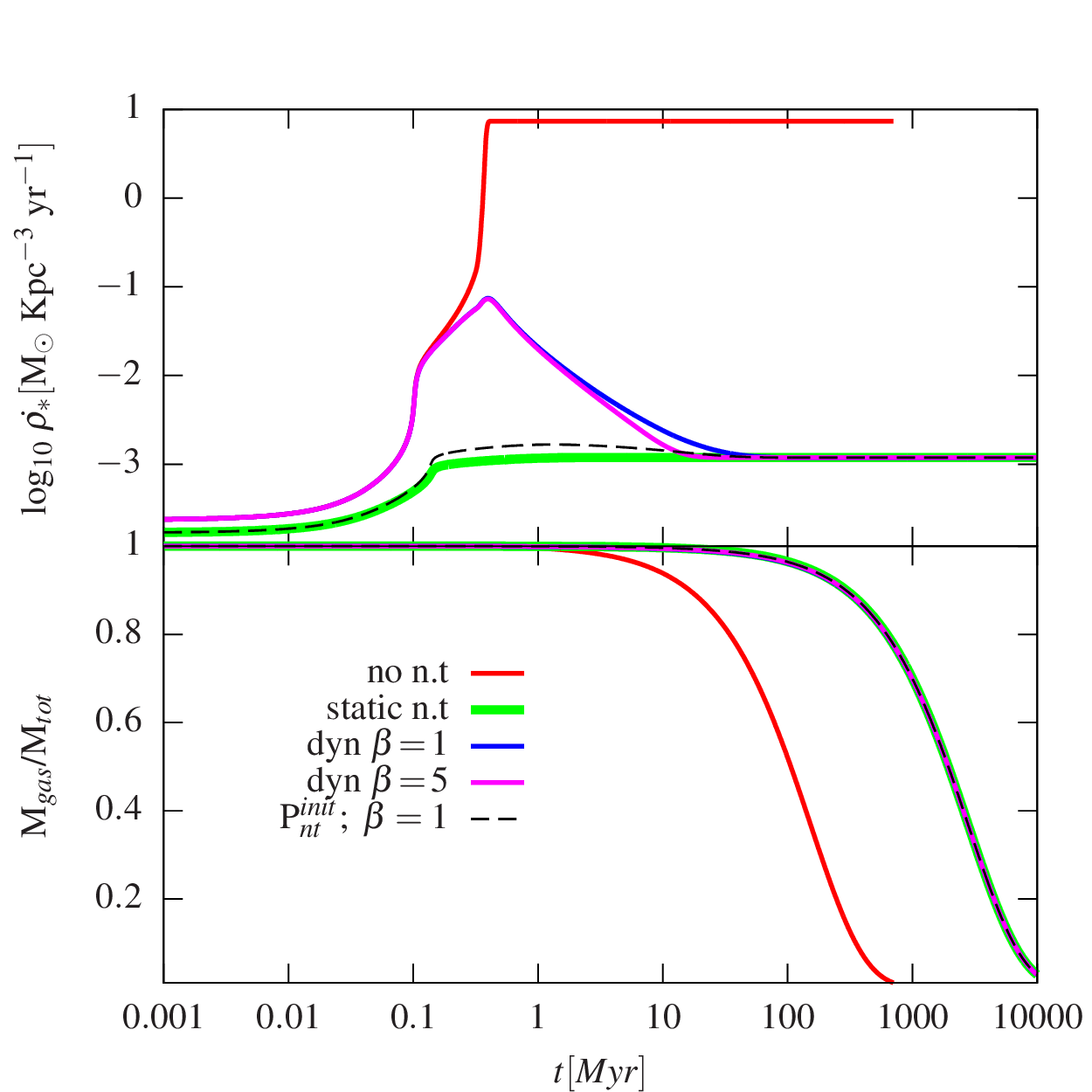}
    \caption{\label{fig:ism_powerlaw2} Same as \Fig{ism_rhodot},
      except that the non-thermal pressure is calculated with Equation
      (\equ{powerlaw2}). 
    }
  \end{center}
\end{figure}

\begin{figure}
\begin{center}
  \includegraphics[width=3.3in]{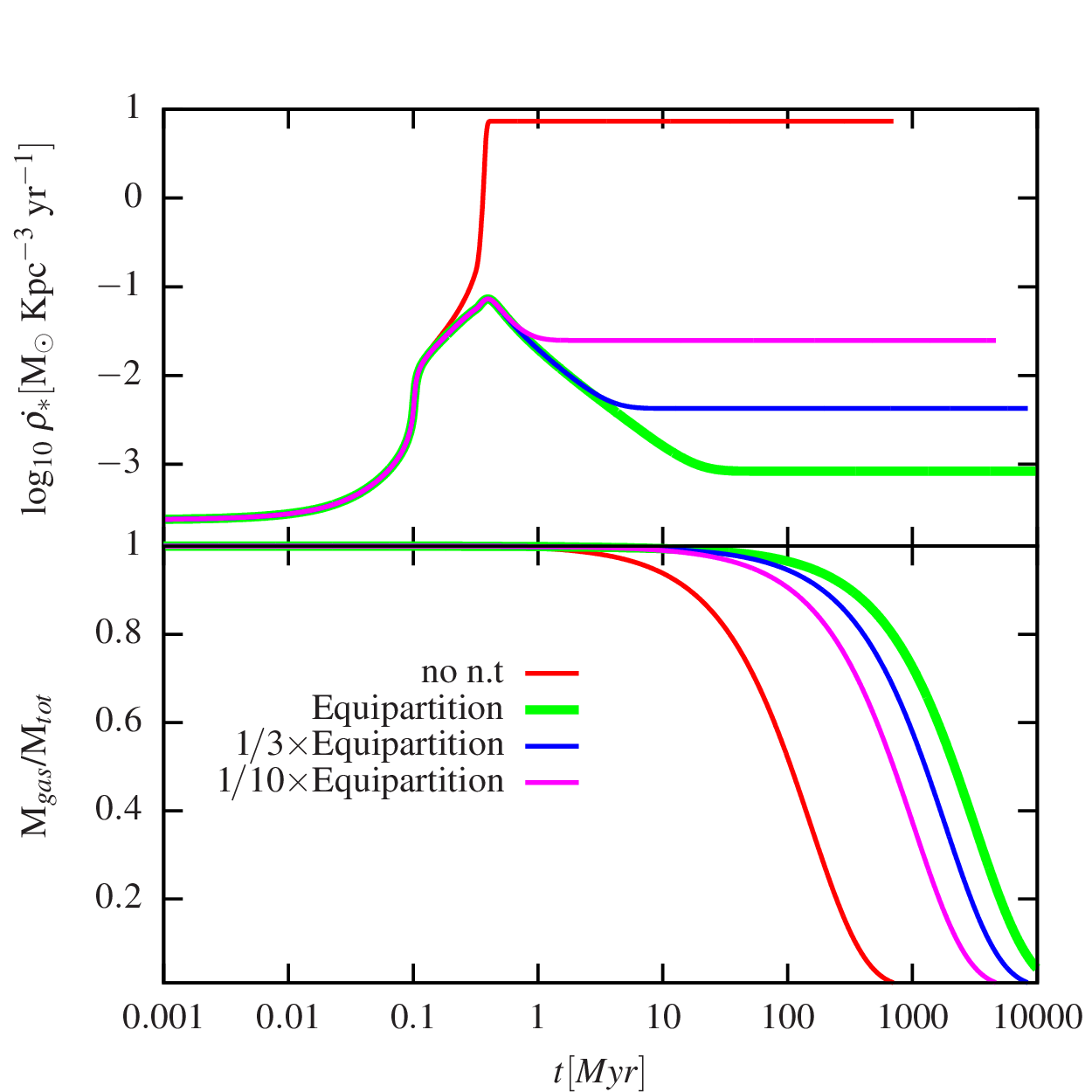}
  \caption{\label{fig:ism_equipartition} Time evolution of the star
    formation rates and the depletion of gas into stars
    ($M_\text{gas}/M_\text{tot}$) for various magnitudes of the non-thermal
    pressure based on the fit in \equ{powerlaw1}) normalized to the
    full equipartition magnitude}
\end{center}
\end{figure}

In order to examine the robustness of the effects of the non-thermal EoS we
repeat the calculations described here for (i) the parameters in
\equ{powerlaw2} and (ii) for a non-thermal component weaker by a
factor of $3$ and $10$; 
\begin{equation}
P_\text{nt}=5.3\times 10^{-13}\left(\frac{\rho}{10^{-24}\text{gr}~\text{cm}^{-3}}\right) \text{erg}~\text{cm}^{-3},
\label{eq:powerlaw3}
\end{equation}
instead of \equ{powerlaw1}. The results are described in
\fig{ism_powerlaw2} and \fig{ism_equipartition}, respectively.

In contrast with the previous calculations, setting $\alpha\ne 1$ in
\equ{powerlaw2} causes the dynamic equation of state to slightly
deviate from the static equation of state even when both calculations
begin with initial conditions of $P_\text{nt}^\text{init}=P_\text{nt}^\text{0}.$ The actual
difference in the evolution between these two cases is still small
in comparison with the difference between them and calculations which begin
with $P_\text{nt}^\text{init}=0.$, for which the gas density overshoots
significantly, and peaks after about 0.5Myr (the dynamic calculation
with a finite initial non-thermal pressure, shown in gray, does over
shoot with the same time scale, but at a much smaller amplitude).

Finally, \fig{ism_equipartition} shows that even when the non-thermal
pressure is reduced by a factor of $10$, the gas depletion time is
still a factor of five or so longer than the depletion time for any of
pure thermal calculations with realistic thermal feedback effeciencies
(see \se{thermal}). Only a perfect thermal efficiency
$\epsilon_\text{SN}=1.$ allows for a depletion time that is comparbale to
the case when a weak non-thermal component is included. This result
emphasizes that non-thermal pressure is far more efficient in delaying
gas depletion to star formation than enahncing thermal feedback from
supernovae. The time scale for relaxation in this weaked non-thermal
pressure case is reduced, however, to one to a few Myr. Our main
conclusion is that any significant non-thermal pressure will
inevitably lead to a large change in the gas depletion time when
compared to pure-thermal pressure models. We infer that this is a
general consequence of non-thermal pressure, regardless of whether
equipartition is assumed.

\subsection{Evolution with a Cutoff Density for Star Formation}
\label{sec:cutoff}
\begin{figure}
\begin{center}
\includegraphics[width=3.3in]{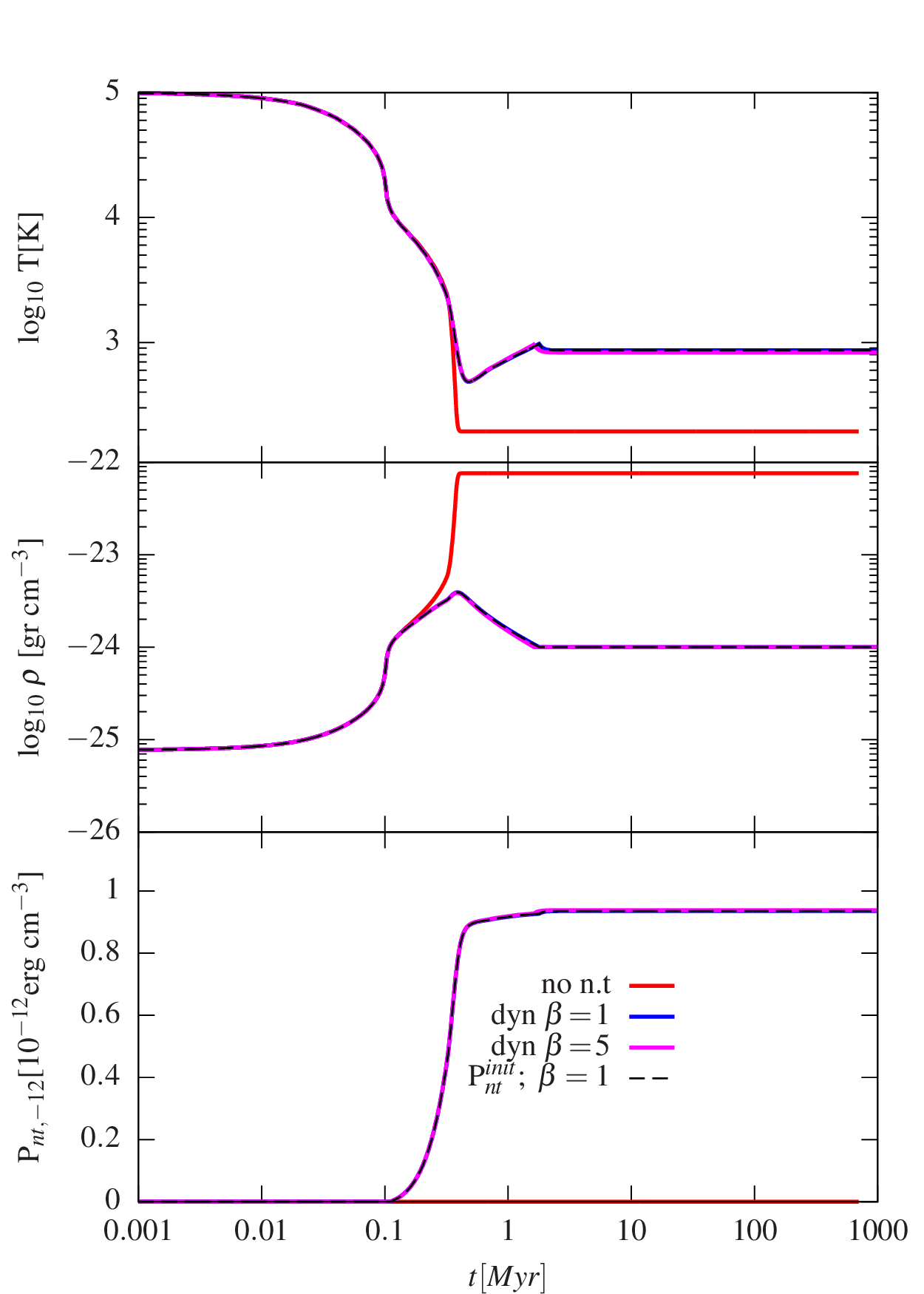}
\caption{\label{fig:ism_sfcutoff} Same as \Fig{ism_trho}, but with
  density dependent star formation cutoff introduced at
  $\rho=10^{-24}\text{gr}\;\text{cm}^{-3}$.  }
\end{center}
\end{figure}

\begin{figure}
\begin{center}
\includegraphics[width=3.3in]{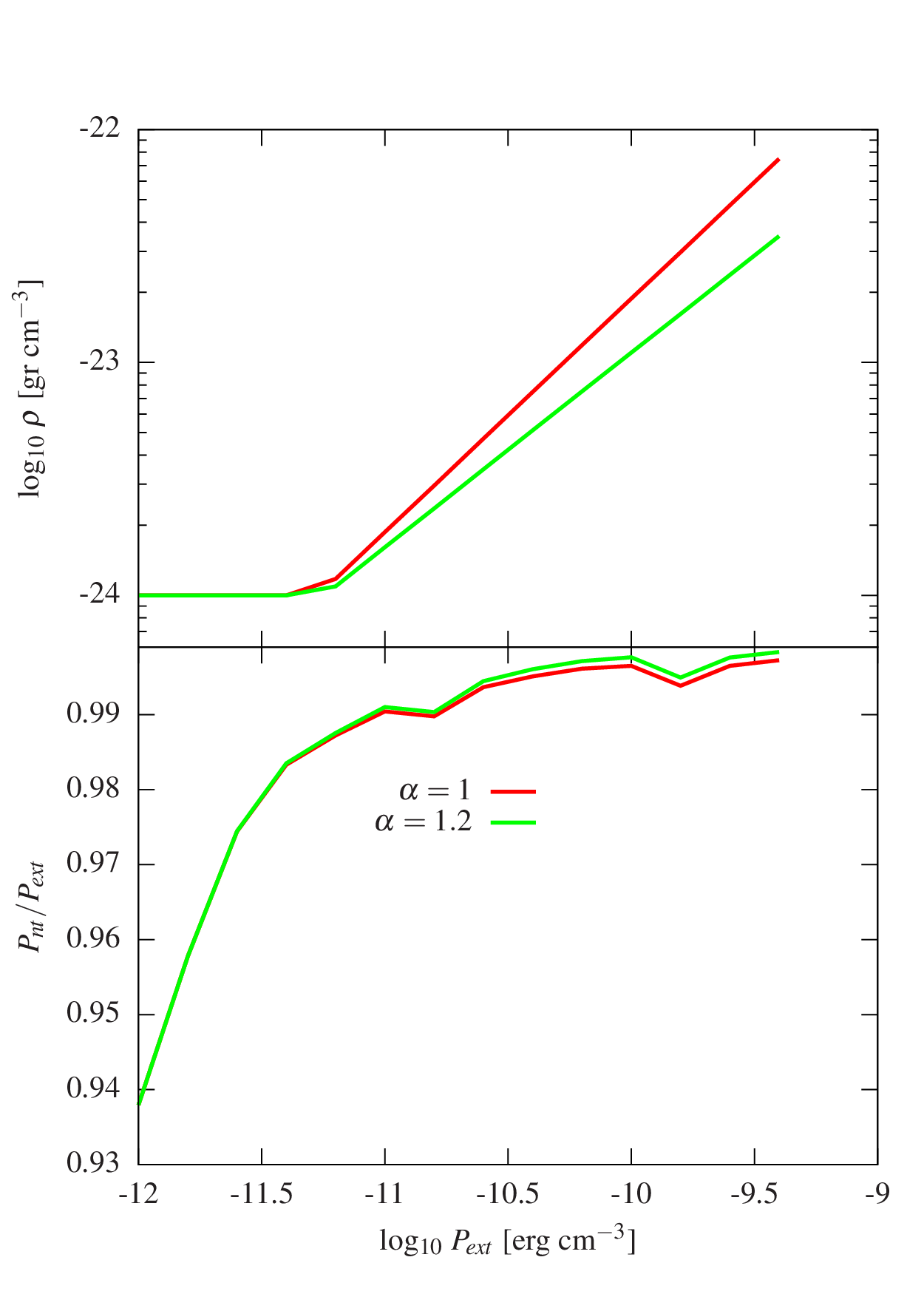}
\caption{\label{fig:extp} The gas density ({\it top panel}) and the
  fraction of the non-thermal pressure from the total pressure ({\it
    bottom panel}) as a function of the external pressure boundary
  condition for the parameters described in table
  \ref{tab:values}. Below $10^-{11.4}~\text{erg}/\text{cm}^3$ the density is always
  the cutoff density for star formation ($10^{-24}\text{gr}/\text{cm}^3$
  here). Above this pressure the density increases as a power-law.  }
\end{center}
\end{figure}

The existence of a star formation density threshold is predicted by
\citet{Kennicutt83} and is present in essentially all numerical models
of galaxy formation. It is typically implemented by invoking a single
numerical value, set to compensate for the inability to simulate star
formation and reach the necessary (high) densities. Moreover, the
thershold is applied to prevent spurious star formation outside of
galaxies. In this subsection we study the response of our model to an
inclusion of such a fiducial threshold. We use a threshold value of
$10^{-24}\text{gr}\;\text{cm}^{-3},$ which is typical in cosmological
simulations. We emphasize that this is a qualitative demonstration of the
effect which is included self-consistently in any numerical
simulation, such as those described below, in \se{ramses}.

\Fig{ism_sfcutoff} describes the temporal evolution of a gas parcel
for a star formation law that includes a sharp cutoff for densities
below $10^{-24}\text{gr}\;\text{cm}^{-3}.$ This threshold is higher than the
equilibrium density calculated without the threshold (\fig{ism_trho})
so there is no strict equilibrium solution (i.e. static solution) for
this case in which the cooling balances the heating at all
times. Instead, we find that the qualitative behaviour of the system is
such, that the response time of the non-thermal components creates a
cycle in which star formation flickers on and off and the
time-averaged heating balances the continuous cooling. By
construction, the single cell model is clearly inadequate for a
quantitative study of this duty cycle, because it coarse grains over
the relevant spatial and temporal scales necessary . We do confirm
numerically that our integration does indeed flicker.

It is noteworthy that applying the density cutoff does not imply that the star formation in a
full simulation in a galactic ISM will occur at constant
density. Non-homogeneity in the ISM is expected \citep[see for example
][]{Ostriker2001} and implies that the external pressure boundary
conditions should vary in space and time. We demonstrate that this principle applies also in the case of non-thermal pressure by
examining the dependence of the equilibrium density and of the star
formation rate on the external pressure conditions. We vary the latter
and solve the equlibrium density and non-thermal pressure for the
non-thermal relations in \equ{powerlaw1} and \equ{powerlaw2}. Our
results are presented in \fig{extp}. We find that the density cutoff
imposes a transition that depends on the external pressures: for low
external pressures, the equilibrium density settles at the cutoff
density for star formation as described above. However, for
$P_\text{ext}\geq 10^{-11.4}~\text{erg/cm}^3$ star formation becomes possible
and the equilibrium density is larger (see the top panel of
\fig{extp}). We note that the non-thermal pressure dominates for
practically any external pressure above this tansition value (lower
panel in \fig{extp}), so the density in this regime essentially scales
as $\rho\propto P_\text{ext}^{1/\alpha}$.  It is encouraging to
note that for pressures that correspond to the plane of the Galaxy
($P\approx 3 \times 10^{12}\text{erg}/\text{cm}^3$) the relative contribution of the
thermal component is a few per-cent,which is in agreement with
observations \citep[see fig. 2 of][and accompanying text]{cox05_ism}.

\section{The effect of non-thermal ISM equation of state on realistic galaxies}
\label{sec:ramses}
We test our model by implementing it on isolated spiral galaxy
simulations ran on \ramses \citep{Teyssier2002}.  In the following
section we will describe in some detail the non-trivial aspects of our
implementation (\se{implementation}), the simulations that were ran
(\se{ramses_ic}), and describe the effects of non-thermal feedback on
the star formation history and on the morphology of the resulting
galaxies (\se{ramses_results}).

\subsection{Model implementation}
\label{sec:implementation}
We now describe the numerical methods we have used to solve for the
Euler equations in presence of a non-thermal energy components. The
original equations have to be modified by adding to the total fluid
energy the non-thermal energy and to the total pressure the
non-thermal pressure. The modified equations now read
\begin{equation}
\frac{\partial \rho}{\partial t} + \nabla \cdot \left( \rho {\bf u}\right) = 0 
\end{equation}
\begin{equation}
\frac{\partial }{\partial t} \left( \rho {\bf u}\right) + \nabla \cdot \left( \rho {\bf u} \otimes {\bf u} + P_\text{tot} {\mathbb I}\right) = - \rho \nabla \Phi
\end{equation}
\begin{equation}
\frac{\partial E_\text{tot}}{\partial t} + \nabla \cdot \left(  {\bf u} \left( E_\text{tot} + P_\text{tot }\right)\right) = - \rho {\bf u}\cdot \nabla \Phi
\end{equation}
\begin{equation}
\frac{\partial e_\text{nt}}{\partial t} + \nabla \cdot \left(  {\bf u} e_\text{nt} \right)  + P_\text{nt } \nabla \cdot {\bf u} = 0 
\end{equation}
The total fluid energy is now defined as
\begin{equation}
E_\text{tot}=\frac{1}{2}\rho u^2 + e + e_\text{nt}
\end{equation}
and the total fluid pressure as
\begin{equation}
P_\text{tot}=P + P_\text{nt}
\end{equation}
where the thermal pressure is given by the EoS of the thermal
component
\begin{equation}
P=(\gamma-1)e,
\end{equation}
and the non-thermal pressure by the EoS of the non-thermal component
\begin{equation}
P_\text{nt}=(\gamma_\text{nt}-1)e_\text{nt},
\end{equation}
For the thermal component only, we can also define the specific
thermal energy $\epsilon$ as
\begin{equation}
e=\rho \epsilon.
\end{equation}
We see in the previous equation that the internal energy of the thermal component is obtained by subtracting from the
total energy the other energy components, namely
\begin{equation}
e=E_\text{tot}- e_\text{nt}-\frac{1}{2}\rho u^2.
\end{equation}
In the hydrodynamics solver, we have to modify several components of
the code to add this non-thermal energy variable.  First, the
predictor step in our MUSCL scheme \citep{Teyssier2002,Fromang2006}
 is augmented by an additional equation for the non-thermal
pressure. Second, in the same predictor step, the non-thermal pressure
is added to the thermal pressure in the equation governing the
velocity update.  The next important correction is for the Riemann
solver, used to define the flux at cell interface, as a function of
the left and right states interpolated with in space and time at the
interface position. We have modified our various approximate Riemann
solver by just replacing the fluid energy and the fluid pressure by
the total energy and the total pressure. The sound-speed in the
augmented hyperbolic system of (quasi-) conservation laws has to be
modified as
\begin{equation}
c^2_\text{s,tot}=\frac{\gamma P + \gamma_\text{nt}P_\text{nt}}{\rho}
\end{equation}
We have tested successfully our new algorithm on simple shock tubes
featuring the additional non-thermal energy.  An important point we
would like to stress is that in our simple model, shock heating occurs
only for the thermal component.  Since in the previous set of
equations, there is no sources of non-thermal energy at shock fronts,
and no coupling between the two energies, the evolution of the
non-thermal component is strictly adiabatic.

For our non-thermal pressure model described in this paper we allow
for one additional scalar component of non-thermal energy with
$\gamma_\text{nt}=2,$ and add a source term for it to the feedback routine
according to the heating portion of \equ{P_nt_dot} and a sink term to
the baryonic cooling routine according to the cooling portion of that
same equation.

\subsection{Simulation parameters}
\label{sec:ramses_ic}
Feedback in hydrodynamical simulations is typically ineffective in
regulating star formation and reducing gas depletion times. To
overcome that, a combination of methods, all ``pumped up'' to be as
efficient as physically possible, is used. We apply here the
'standard' tools used in the \ramses runs of the AGORA \citep{kim2014}
isolated galaxy.  Our base-line simulation, ``Standard sim'' uses the
standard tools used in RAMSES which include several ad-hoc measures
calibrated to prevent reaching high density and short depletion
times. These methods include delayed cooling (preventing cooling for a
period of time after feedback energy injection to account for the
adiabatic phase of the Sedov-Taylor explosion). Additional methods
applied is to increase the stochasticity of the process by allowing
feedback to operate according to a Poisson distribution with a typical
mass scale of a giant molecular cloud (GMC) and, although not strictly
motivated by feedback, incorporating a pressure floor for the ISM gas
that prevents it from reaching extremely cold and dense states which
would imply very large star formation rates. We aim here to
demonstrate that the parameters of these methods can be relaxed once
our non-thermal feedback model is used. For comparison we
consider two more simulations. In the first, delayed cooling is turned
off and the simulation indeed exhibits an over-production of stars. In
the third simulation we introduce our non-thermal feedback and
demonstrate its ability to reduce star formation without the delayed
cooling model.

The three simulations are defined in table \ref{tab:sims}. The
simulations were ran using the AGORA low resolution initial conditions
for a Milky way-like galaxy. The setup consists of a DM halo of
$10^{12}\msun$, stellar disk of $3.4\times 10^{10}\msun$ and a gaseous
disk of $8.5\times 10^9\msun.$ The maximal refinement level is $12$,
with a box size of $400\kpc$ and a maximal resolution of $\sim
200\pc.$ Particle mass was $3\times 10^5\msun$ for the stellar
component and $10^7\msun$ for the halo dark matter
particles. Execution time until a physical time of $2\Gyr$ on $240$
cores on the ICPL\footnote{\url{http://icpl.huji.ac.il}} cluster took
a few days.

In all simulation the GMC mass is $6.4\times 10^6\msun,$ and the
pressure floor is defined as a minimal temperature of
$T_\text{min}=10^4(n/0.1)^{2/3}$\,K. For the 'standard feedback' simulation
the delayed cooling timescale is $20\Myr.$ For the non-thermal
feedback simulation the amount of energy that is injected into the
non-thermal component is half of the energy that is injected into the
thermal part, the asymptotic value for the non-thermal component was
set by \equ{powerlaw1} and the rate of energy dissipation was set to
$\beta=1/4$ (\equ{cooling1}).

\begin{table}
\caption{Description of simulations.}
\small
\begin{tabular}{ r r r }
\hline
\hline
Name & delayed cooling & non-thermal pressure\\
\cline{1-3}\\
Standard sim&  on & off \\ 
Weak sim    & off & off \\ 
Non-thermal sim  & off & on \\ 
\hline
\hline
\end{tabular}
  \label{tab:sims}
\end{table}

\begin{figure}
\begin{center}
\includegraphics[width=3.3in]{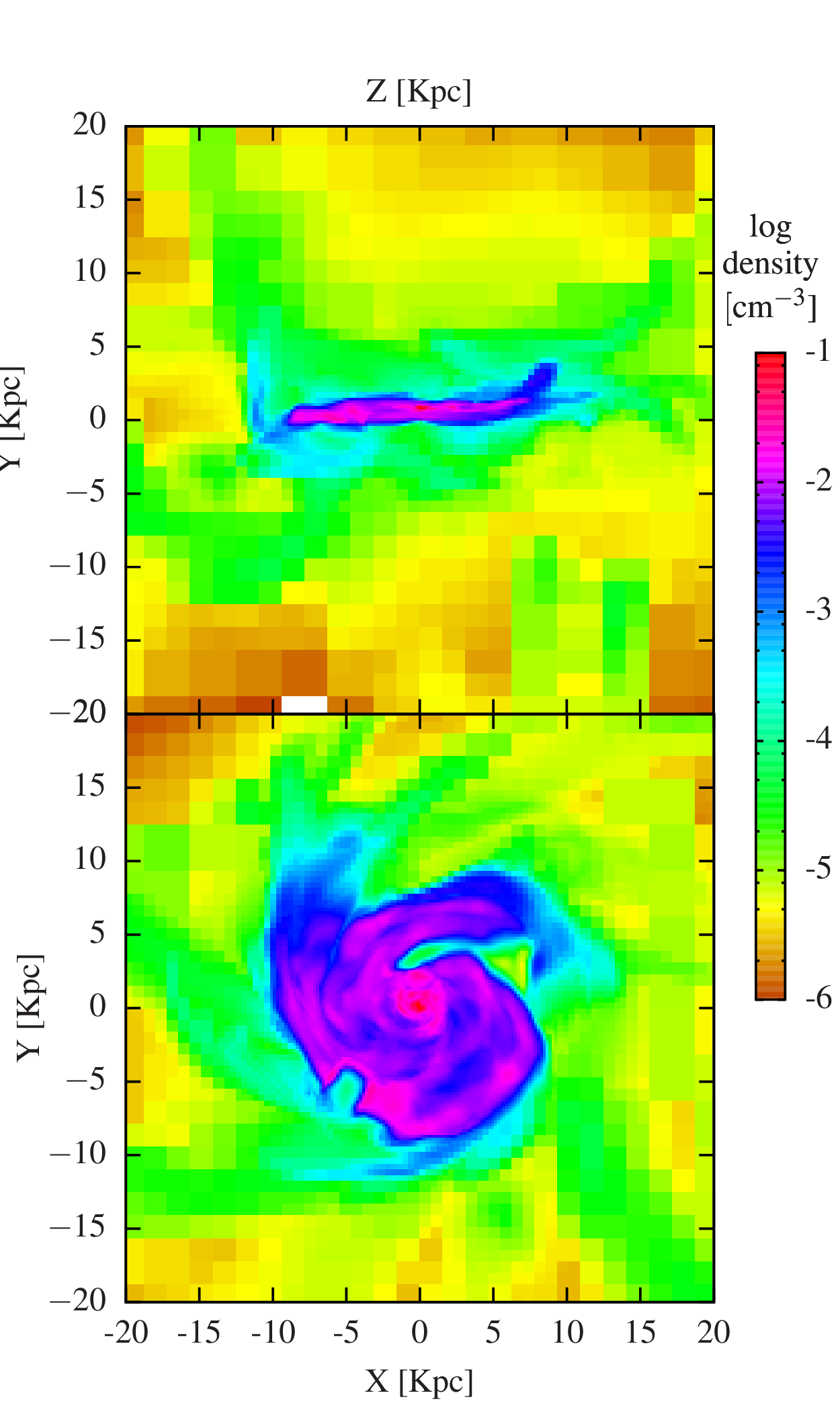}
\caption{\label{fig:mw2a} Density colormap of the central $40\kpc$ of
  our ``Standard sim'' simulation at $t=1100\Myr$. {\it Top panel}
  edge on, {\it bottom panel} face on. }
\end{center}
\end{figure}

\begin{figure}
\begin{center}
\includegraphics[width=3.3in]{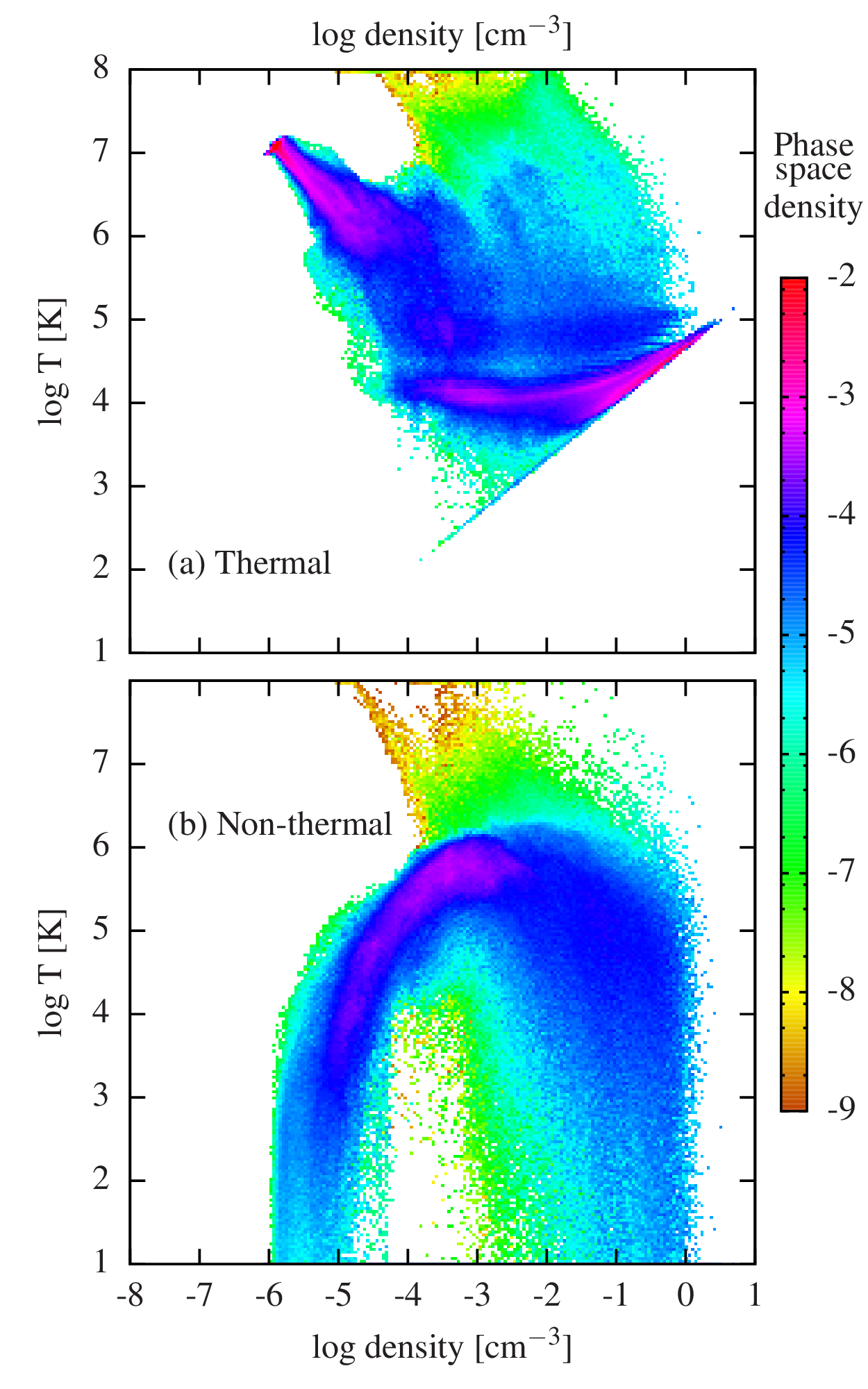}
\caption{\label{fig:histnt} Density-Temperature histograms of
  ``Non-thermal sim''. {\it Panel (a)} presents the thermal component,
  and {\it panel (b)} the non-thermal component at time $t=1100\Myr.$
  The temperature of the 'non-thermal' component is defined as the
  non-thermal pressure divided by the density.}
\end{center}
\end{figure}

\begin{figure}
\begin{center}
\includegraphics[width=3.3in]{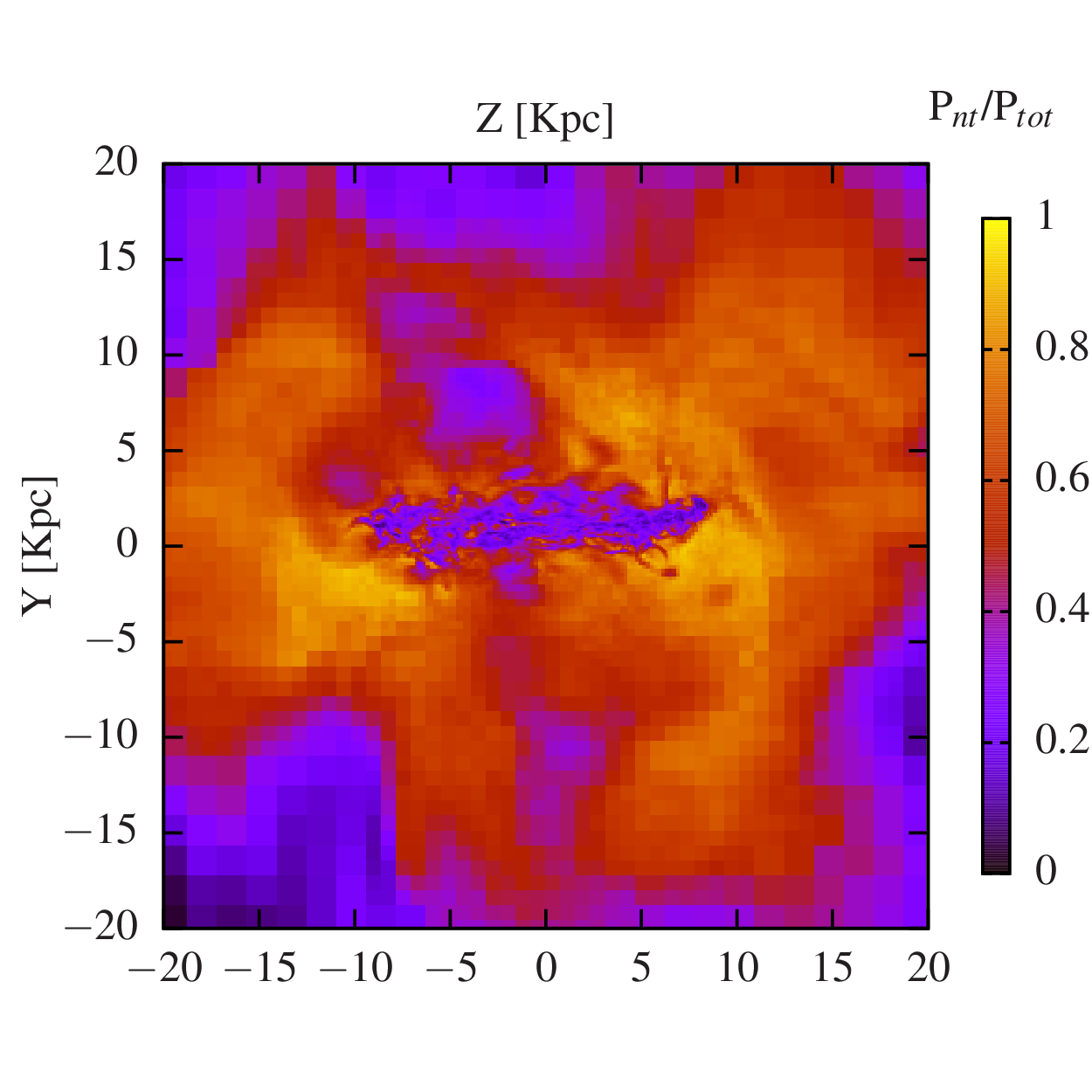}
\caption{\label{fig:ppnt} The spatial extent of the non-thermal
  pressure at $t=1100\Myr.$ The colormap represents the ratio between the
  non-thermal pressure to the total pressure.}
\end{center}
\end{figure}
\subsection{Results}
\label{sec:ramses_results}
\begin{figure}
\begin{center}
\includegraphics[width=3.3in]{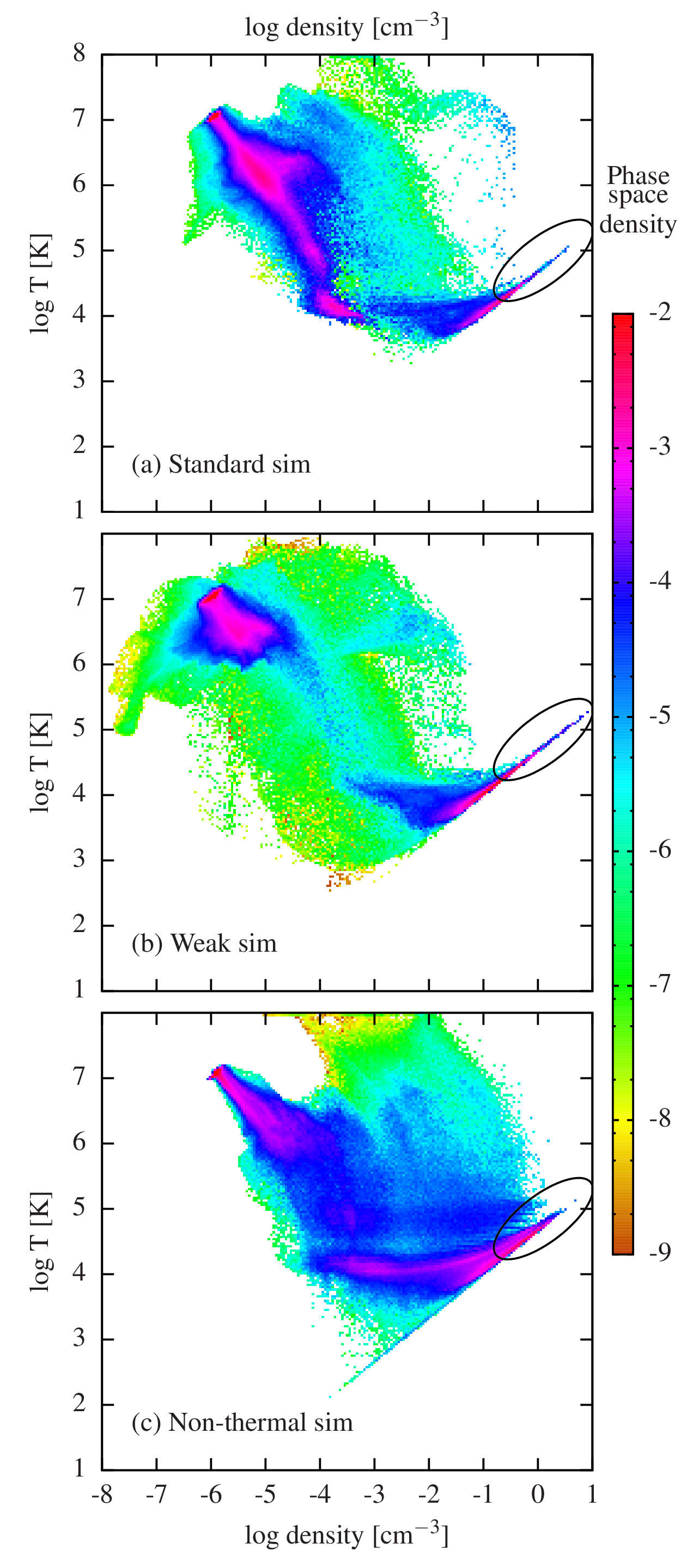}
\caption{\label{fig:hists} Density-Temperature histograms of the three
  models for feedback at time $t=1100\Myr.$ {\it Panels (a), (b) and
    (c) } show results from ``Standard sim'', ``Weak sim'' and
  ``Non-thermal sim'' respectively. The black ellipses point to the
  star forming region on the n-T plot.}
\end{center}
\end{figure}

\begin{figure}
\begin{center}
\includegraphics[width=3.3in]{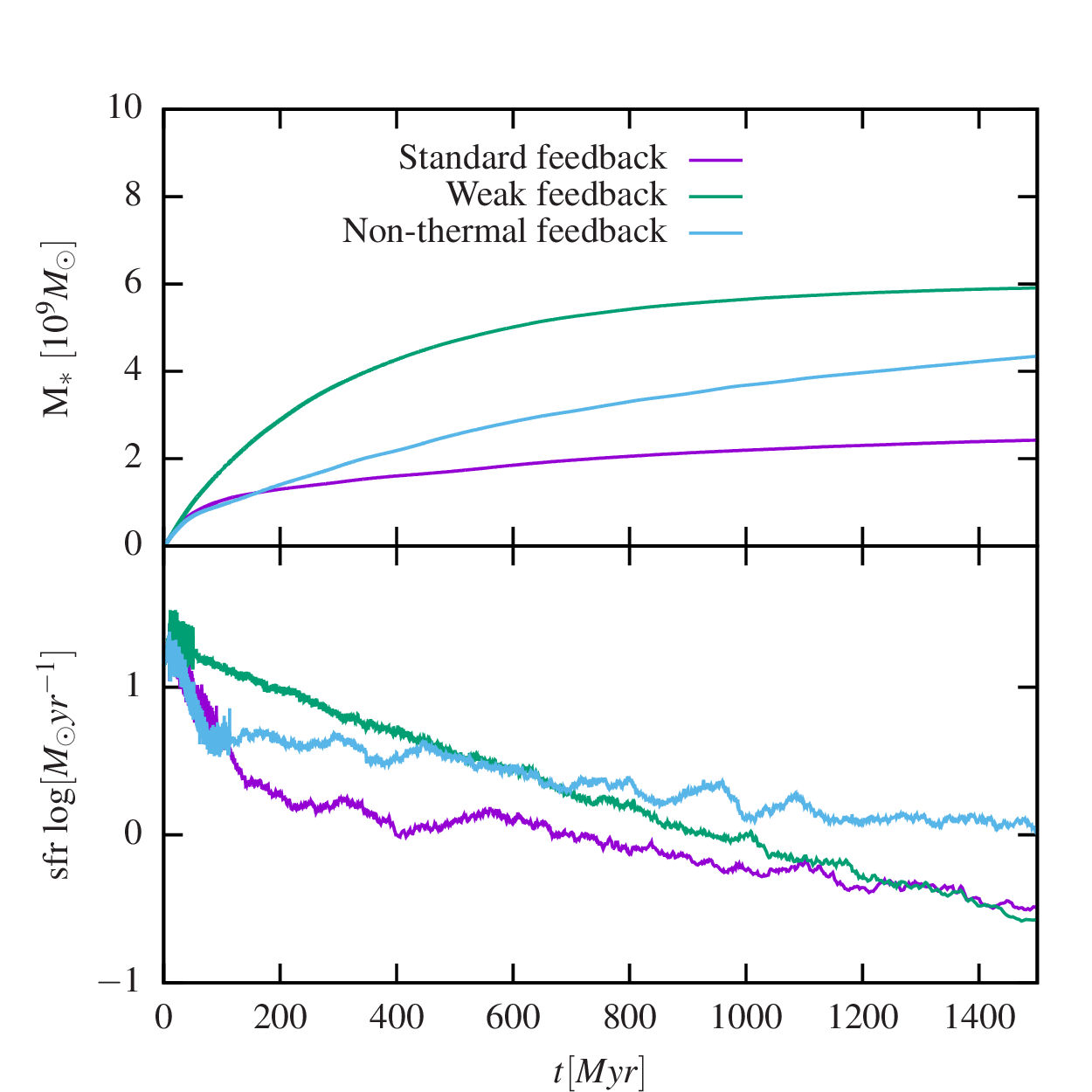}
\caption{\label{fig:sfrcomp} Star formation histories ({\it top
    panel}) and star formation rates ({\it bottom panel}) for
  simulations with the three feedback models.}
\end{center}
\end{figure}

Some results of the simulations described in \se{ramses_ic} are
presented in \fig{mw2a}-\fig{sfrcomp}. \fig{mw2a} shows a head-on and
edge-on view of ``Standard sim''. The simulation produces a stable
rotating disc, and strong outflowing winds that are noticeable as the
rough ``x'' shaped overdensity extending diagonally to the edges of our
$40\kpc$ box. We note that in all these simulations there is no hot
halo gas component, that is expected to strongly affect outflows.

\fig{histnt} presents a density-temperature histogram of our
``Non-thermal sim''. The top and bottom panels show the thermal and
non-thermal components. For the non-thermal component we define the
effective temperature as the non-thermal pressure divided by the
density. In the histograms, isobaric lines would be represented by
lines with slope of $-1$ ($P\sim n \times T=\textrm{const}$) and the
pressure floor is the sharp cutoff diagonal through the lower right
part of the plot with slope of $2/3.$ The non-thermal temperature is
typically higher than the thermal pressure. For the formulation of the
non-thermal pressure used here (\equ{powerlaw1}), this asymptotic
temperature is independent of density and is simply $P/\rho
m_\text{p}/k_\text{B}=6.4\times 10^4$\,K, with $m_\text{p}$ the proton mass and $k_\text{B}$
Boltzmann's constant. For high densities, when the non-thermal cooling
is efficient enough for gas to relax into the asymptotic value, this
value roughly coincides with the effective temperature for the
non-thermal component. However, the scatter due to the stochasticity
of the feedback process is very large.

\fig{ppnt} shows an edge-on view of the galaxy with the colormap
representing the fraction of the non-thermal pressure to the total
pressure. It is evident that dynamically significant non-thermal
pressure exists around the disk at a distance of a few kiloparsecs,
consistent with magnetic field observations \citep{ferriere01}.

\fig{hists} compares the thermodynamic state of the gas of the various
simulations. Stars are formed through the high density branch with a
slope of $2/3$, emphasized by the black ellipse in the plots.  While
this line is truncated in ``Standard sim'' ({\it panel (a)}), at
$n\sim 3\text{cm}^{-3},$ is continues beyond $10 \text{cm}^{-3}$ for
``Weak sim'' ({\it panel (b)}) which causes an increase in star
formation (see \fig{sfrcomp}). The third panel is similar to the top
panel of \fig{histnt}. When non-thermal feedback is added, the amount
of high-density gas is reduced. For the non-thermal case ({\it panel
  (c)}) it is evident that there is a considerable fraction of the gas
at intermediate densities ($10^{-4}-10^{-1} \text{cm}^{-3}$) and
$~10^4$\,K temperature that is absent from the two purely thermal
simulations. This is a result of the isochoric cooling process that
occurs in that case (corresponding to purely vertical thermodynamic
trajectories on the histogram plot) versus the more efficient isobaric
cooling for the thermal cases (corresponding to diagonal trajectories
going down and to the right, with slope of $-1$, as explained
above). The extra non-thermal pressure supports the gas and allows it
to cool without contracting to the pressure floor.  In ``Standard
sim'' ({\it panel (a)}), most of the gas has been blown out of the
galaxy to $T=10^6-10^7$\,K and $n=10^{-7}-10^{-4}$ which effectively
shut down star formation (see \fig{sfrcomp}). The non-thermal pressure
({\it panel (c)}) affects intermediate densities and keeps the gas
pressurized with pressure corresponding to $\sim 10^5$\,K, without
blowing it out of the galaxy altogether.

\fig{sfrcomp} shows the star formation history and star formation
rates for the three simulations. The most efficient quenching occurs
for ``Standard sim'' and is a result of massive blowout of gas from
the galaxy. ``Weak sim'' depletes most of the gas rapidly, within the
first $\sim 500\Myr$ of the simulation, demonstrating the overcooling
catastrophe that occurs for insufficient feedback. ``Non-thermal sim''
exhibits regulated star formation that produces stars at $\sim
1\msun/\yr$ throughout the simulation.

Not surprisingly, our results indicate that
non-thermal pressure can have a significant effect on star formation
and prolong the depletion times of galaxies to the observed
timescales. Clearly further research will be required to quantify the
relative importance of the non-thermal components to other
feedback mechanisms that are applied. Regardless, we emphasize that
observational evidence, as well as robust theoretical motivation, point
to the existence of this component and that, in one form or another,
it should be incorporated in cosmological and galactic simulations.

\section{Summary and discussion}
\label{sec:summary}
On galactic scales, the interstellar medium exists at quasi-static
pressure that is required to support the atmosphere above it. In an
equilibrium configuration, loss of pressure due to cooling processes
is balanced by heating, which for typical disc galaxies at low
redshifts, is dominated by stellar feedback.  Stellar feedback,
through its dependence on star formation rate is related to the ISM
gas density.

It is a well-known result that when only thermal pressure is
considered in simulations, the resulting ISM density constrained by
the pressure and heating-cooling equilibria leads to relatively large
star formation rates and short ($\sim 100$Myr) gas depletion time
scales. This is considerably faster than depletion times of $\approx
1\Gyr$ inferred from observations
\citep{Kong2004,Bauermeister2013,Pflamm-Altenburg2009,Tacconi2013}. In
this work we consider the contribution of non-thermal pressure
components to this picture. Non-thermal pressure consists of
turbulence, cosmic rays and magnetic fields, and we examine their
impact in an effective model. Current cosmological simulations generally
do not include the later two, and do not always resolve
turbulence. We demonstrate that non-thermal pressure
components can be instrumental in solving the depletion time discrepancy
in two respects: they reduce the quasi-steady state density and the
corresponding star formation rates and cooling times, and they
stabilize the gas by adding longer relaxation times in cases where
star formation flickers on and off. The regulating effect has been
shown previously for cosmic rays
\citep{Salem2014,Booth2013,Hanasz2013} and turbulence
\citep{Ostriker2001,Braun2012} but the two were not considered
together and in any case were not yet formulated in a way which is
applicable to large scale cosmological simulations.

To test our assumptions we construct a simplified physical model for
which all the non-thermal components achieve a steady state that is
solely a function of density. While simplistic, the advantage of such
an approach is that it is readily applicable in numerical
simulations. Furthermore, we calibrate this density dependence by
using the observed relations between the star formation rate for
various galaxy observations and the synchrotron radiation, so that the
magnitude of the effect is reasonably constrained.  To study its
effect we first implement it into a single-zone numerical model that
traces the evolution of a parcel of star forming gas with varying
physical conditions under isobaric boundary conditions that mimic the
pressure confinement of the gas by the atmosphere around and above
it. Using this mode we find that for a given, realistic, thermal
feedback the depletion times naturally grow from $\approx 100\Myr$ to
$\approx 2\Gyr$ in better agreement with observations, and that the
coarse grained density of the gas is reduced by several orders of
magnitude. Then, the model is implemented into the hydrodynamical code
\ramses and we present three simulations of the same isolated
Milky-way-like galaxy with three different physical models. In the
first we use some of the ``standard'' recipes generally used for
feedback. Using that model reduces star formation by blowing the gas
to high temperatures and low densities, and expelling it from the
galaxy. Then, to demonstrate the problem we deliberately turn off one
of the key feedback components - the delayed cooling and show that gas
cools and accumulates at the numerical pressure floor with high
densities that cause a depletion of the stars within $\sim 100\Myr$ -
at odds with observed depletion times of $1-2\Gyr.$ In the third
simulation we introduce the non-thermal model calibrated by the
observed radio FIR relation. For that model, the gas remains
pressurized at intermediate temperatures and densities, reducing the
non-physical low density gas of the first model, and the non-physical
high density star forming gas of the second model. This model is
effective in regulating star formation for a long period of time
($\sim 1\Gyr$) without blowing the gas out of the galaxy altogether.

This work is a natural first step in incorporating non-thermal
pressure components in galactic scale simulation. The next steps can,
and should, pursue several avenues of research. The first is to better
model the various non-thermal components, their internal interaction
and their interaction with the thermal component and star
formation. This would relax the assumption of equipartition, and
replace the observational constraints with more physically motivated
ones. For this step, calibration against results from ISM-scale
hydrodynamic simulations will be beneficial. A different avenue to
pursue, in tandem or separately, is to use our \ramses patch to run
large scale numerical simulations and to demonstrate its effect and
applicability on cosmic scales. Cosmological simulations today
generally do not include the magnetic fields and cosmic rays, and do
not always resolve turbulence, and our approach allows to circumvent
this difficulty by using a simple effective parametrization. Such
simulations will naturally include realistic boundary conditions for
the ISM, namely the halo gas, and allow us to study its interactions
with the ISM and its effect on winds. Ultimately, once large scale
cosmological simulations are possible with all the necessary physics,
a toy sub-grid model can also be calibrated directly to those
simulations and used as a cheaper approximation for them.

\section*{Acknowledgements}
We thank Andrey Kravtsov for making the cooling tables available to us for the single cell calculations. Computational resources were provided through ICPL (\url{http://ICPL.HUJI.AC.IL}).



\label{lastpage}

\end{document}